\documentclass[traditabstract,fleqneferee]{aa}

\usepackage{amsmath}
\usepackage{txfonts}
\usepackage{graphicx}
\usepackage{natbib}
\bibpunct{(}{)}{;}{a}{}{,}

\begin{document}

\newcommand{\be}{\begin{equation}}
\newcommand{\ee}{\end{equation}}
\newcommand{\bea}{\begin{eqnarray}}
\newcommand{\eea}{\end{eqnarray}}
\newcommand{\dev}{D}
\newcommand{\der}{\nabla}
\newcommand{\Atil}{{\tilde A}}
\newcommand{\Ahat}{{\hat A}}
\newcommand{\Ebar}{{\hat E}}
\newcommand{\Sbar}{{\hat S}}

\title{GRMHD in axisymmetric dynamical spacetimes: the X-ECHO code}

\author{N. Bucciantini    \inst{1}
\and      L. Del Zanna      \inst{2}
}

\institute{
NORDITA, AlbaNova University Center, Roslagstullsbacken 23, 10691Stockholm, Sweden
\\ \email{niccolo@nordita.org}
\and
Dipartimento di Fisica e Astronomia,
Universit\`a di Firenze, Largo E. Fermi 2, 50125 Firenze, Italy
\\ \email{luca.delzanna@unifi.it}
}

\date{Received... ; accepted...}

\abstract
{We present a new numerical code, X-ECHO, for general relativistic 
magnetohydrodynamics (GRMHD) in dynamical spacetimes.
This is aimed at studying astrophysical situations where strong
gravity and magnetic fields are both supposed to play an important
role, such as for the evolution of magnetized neutron stars or for the  
gravitational collapse of the magnetized rotating cores of massive stars, 
which is the astrophysical scenario believed to eventually lead to (long) GRB events. 
The code is based on the extension of the \emph{Eulerian conservative high-order} 
(ECHO) scheme [\emph{Del Zanna et al., A\&A 473, 11 (2007)}] for  GRMHD, 
here  coupled to a novel solver for the Einstein equations in the extended 
conformally flat condition (XCFC). 
 We solve the equations in the $3+1$ formalism, assuming axisymmetry
and adopting spherical coordinates for the conformal background metric.
The GRMHD conservation laws are solved by means of shock-capturing methods 
within a finite-difference discretization, whereas, on the same numerical grid, 
the Einstein elliptic equations are treated by resorting to spherical harmonics 
decomposition and solved, for each harmonic, by inverting band diagonal matrices.
As a side product, we build and make available to the community a code
to produce GRMHD axisymmetric equilibria for polytropic relativistic stars 
in the presence of differential rotation and a purely toroidal magnetic field.
This uses the same XCFC metric solver of the main code and has been
named XNS. Both XNS and the full X-ECHO codes are 
validated through several tests of astrophysical interest.
}


\maketitle


\section{Introduction}
\label{sect:intro}

The most spectacular phenomena in high-energy Astrophysics,
like those associated to active galactic nuclei (AGNs), 
galactic X-ray binary systems, or gamma-ray bursts (GRBs),
typically involve the presence of rotating compact objects and magnetic fields.
In some cases, such as the merging of binary systems (formed by either
neutron stars, NSs, or black holes, BHs) or the collapse of rotating cores of
massive stars towards a NS or BH, the interplay between matter, 
electromagnetic fields, and gravity is so strong that the MHD equations
governing the fluid motions must be solved self-consistently with the
Einstein equations for the spacetime metric. Even for less violent
phenomena, as the oscillations of neutron stars \citep{font02}, a self
consistent solution of the fluid equations together with the
spacetime evolution is essential to properly estimates the
frequencies of the eigenmodes.
In the case of a binary NS merger, which is a possible mechanism to
account for short GRBs, a strong magnetic field could be produced
due to the induced shear \citep{price06}. Long GRBs are instead associated
to supernova events and to the core collapse of massive stars \citep{woosley06},
leading to the subsequent formation of a rotating and strongly magnetized
compact object. The mainstream \emph{collapsar} model 
\citep{woosley93} implies the rapid formation of a maximally rotating 
Kerr BH at the center, accreting material from a torus and likely to
loose energy through the Blandford-Znajeck mechanism \citep{barkov08}. 
However, a promising alternative for the GRB central engine involves the 
presence of a millisecond magnetar with $B\gtrsim 10^{15}$~G \citep{usov92}.
The origin of such enormous fields for magnetars is probably due
to efficient dynamo action during the neutrino cooling phase
in the hot, deleptonizing proto-NS \citep{duncan92}.
On the observational side, magnetars are the accepted explanation
for anomalous X-ray pulsars and soft gamma-ray repeaters
\citep{kouveliotou99}.

On the computational side, the last decade has witnessed a very
rapid evolution in the construction of shock-capturing codes for 
general relativistic MHD (GRMHD) in both static and dynamical spacetimes,
with a wealth of astrophysical applications to the situations outlined above
\citep{font08}. In the present paper we describe a novel code for GRMHD
in dynamical spacetimes, named X-ECHO, aimed
at studying the evolution of magnetized relativistic stars and the  
gravitational collapse of the magnetized rotating cores of massive stars.
X-ECHO is built on top of the \emph{Eulerian conservative
high-order} code \citep{delzanna07} for GRMHD in a given
and stationary background metric (Cowling approximation),
which in turn upgraded the previous version for a Minkowskian
spacetime \citep{delzanna02, delzanna03}. 
ECHO relies on robust shock-capturing methods within a finite-difference
discretization scheme (two-wave Riemann solvers and limited high order
reconstruction routines), with a staggered constrained transport
method to preserve the divergence-free condition for the magnetic
field (to machine accuracy for second order of spatial accuracy),
as proposed by \cite{londrillo00, londrillo04}. 
The ECHO code has been already successfully applied to a variety of 
astrophysical situations involving magnetized plasmas around
compact objects, like the dynamics and non-thermal emission of
pulsar wind nebulae \citep{bucciantini03, delzanna04, bucciantini04a, 
bucciantini05a, bucciantini05b, delzanna06, volpi08}, emission of relativistic 
MHD winds from rotating NSs \citep{bucciantini06a}, magnetar winds producing
long GRB jets escaping the stellar progenitor \citep{bucciantini08,bucciantini09}, 
and post-merger accreting disks around Kerr BHs \citep{zanotti10}.
In spite the code being fully 3D, due to the nature of the sources, 
invariably a plasma surrounding a central compact object,
all the above applications have been performed in 2D axisymmetric 
spacetimes using spherical-type coordinates (either in Minkowski, 
Schwarzschild, or Kerr metric). The X-ECHO version presented and tested
here shares the same philosophy, and, in view of the future applications
mentioned above,  only the axisymmetric case will be considered.

The Einstein and GRMHD equations in X-ECHO are written by fully 
exploiting the so-called $3+1$ formalism (like in ECHO), in which the 
original equations are split in their temporal and spatial components.
The $3+1$ formalism is nowadays adopted in basically all numerical
schemes for general relativity \citep{alcubierre08, baumgarte10}, where the system 
of Einstein equations is treated like a Cauchy problem with some initial data
to be evolved in time through hyperbolic equations. However,
like for the solenoidal condition for the magnetic field, non-evolutionary
constraints must be preserved in the numerical evolution and
computational methods for modern codes are divided into two main classes:
1) \emph{free-evolution} schemes, mainly based on hyperbolic equations alone, 
where this problem is alleviated by appropriate reformulations of the equations
\citep[BSSN:][]{shibata95, baumgarte99}, eventually with the addition of 
propagating modes and damping terms \citep[Z4:][]{bona03, bernuzzi10};
2) \emph{fully constrained} schemes, where the constraints are enforced
at each timestep through the solution of elliptic equations \citep{bonazzola04},
a more robust but computationally demanding option, since elliptic solvers
are notoriously difficult to parallelize. 
Most of the state-of-the-art 3D codes for GRMHD in dynamical spacetimes
are based on free-evolution schemes in Cartesian coordinates
\citep{duez05, shibata05, anderson06, giacomazzo07, montero08, farris08}, 
and have been used, in the case of magnetized plasmas, for gravitational collapse
\citep{duez06a, shibata06, shibata06b, stephens07, stephens08}, evolution of NSs 
\citep{duez06b, kiuchi08b, liebling10}, binary NS mergers 
\citep{anderson08, liu08, giacomazzo09, giacomazzo10}, and
accreting tori around Kerr BHs \citep{montero10}.

Provided the emission of gravitational waves is not of primary interest, 
a good option in the class of fully constrained scheme is represented by the
\emph{conformally flat condition} (CFC) schemes 
\citep[e.g.][]{wilson03, isenberg08}, 
an approximation often employed for the study of gravitational
collapse or NS stability and evolution \citep{dimmelmeier02, 
saijo04, dimmelmeier06, cerda-duran08, abdikamalov09}.
CFC is typically associated to axisymmetric configurations and spherical
coordinates, and it is exact in the spherically symmetric case.
Deviations from full GR solutions in the axisymmetric case for
this kind of applications have already been shown to be negligible 
\citep{shibata04, ott07}, though the nonlinear equations may 
show serious uniqueness problems for highly compact NSs or nascent BHs
\citep[see][and references therein]{cordero-carrion09}.
Moreover, CFC requires the solution of all the elliptic equations for
the metric terms at the same time, usually by means of iteration of Poisson 
solvers, together with the inversion of conservative to primitive
fluid/MHD variables, another iterative numerical process.
All these difficulties have been resolved recently by the \emph{extended
conformally flat condition} (XCFC) formulation \citep{cordero-carrion09}:
all the elliptic equations (now eight rather than five) are hierarchically
decoupled and local uniqueness is ensured \citep[see also][]{saijo04},
thus this will be our choice for X-ECHO. 

In the present work we propose and test a new numerical solver based on 
XCFC for an axisymmetric spacetime in conformally flat spherical-like
coordinates. We employ the same numerical grid used for
the evolution of the fluid and magnetic quantities through
the ECHO scheme, and the Poisson-like equations are solved
through a hybrid method based on spherical harmonics 
decomposition and direct inversion of band diagonal matrices,
resulting from a second order finite difference discretization
of the radial equations, for each harmonic.
As a side product, we build and make available to the community a
numerical code based on XCFC to produce self-consistent GRMHD 
axisymmetric equilibria for polytropic relativistic stars 
in the presence of differential rotation and toroidal magnetic fields,
here named XNS. Both XNS and the full X-ECHO codes are validated 
through several tests of astrophysical interest, including
accuracy checks in the initial data for various NS 
equilibrium configurations, accuracy in finding the frequencies
of their normal modes of oscillations, an evolutionary test of migration
of NS unstable equilibria to stable branches, a test on the stability
of a differentially rotating, magnetized NS with a toroidal field, 1D and 2D
collapse of an unstable NS toward a BH, and a \emph{toy collapse}  
of a differentially rotating NS with poloidal fields, 
as a first step towards more realistic magneto-rotational 
core collapse simulations.

The paper is structured as follows. In Sect.~\ref{sect:3+1} we introduce 
 and review the Einstein equations in the $3+1$  formalism, 
first in their general form and then specialized in the CFC approximation,
and we review the GRMHD equations.
In Sect.~\ref{sect:x-echo} we discuss the new numerical XCFC
solver assuming axisymmetry and spherical coordinates
for the conformal flat 3-metric, whereas the description
of our novel XNS code for NS initial data can be found in Sect.~\ref{sect:xns}.
Numerical validation and testing of various cases of NS equilibria, oscillations 
and collapse are presented in Sect.~\ref{sect:tests}, while
Sect.~\ref{sect:concl} is devoted to the conclusions.
In the following we assume a signature $\{-,+,+,+\}$ for the spacetime metric and 
we use Greek letters  $\mu,\nu,\lambda,\ldots$ (running from 0 to 3) for 4D 
spacetime tensor components, while Latin letters  $i,j,k,\ldots$ (running from 1 to 3) 
will be employed  for 3D spatial tensor components. Moreover, we set 
$c=G=M_{\sun}=1$ and we absorb the $\sqrt{4\pi}$ factors in the definition
of the electromagnetic quantities.


\section{Basic equations in the $3+1$ formalism}
\label{sect:3+1}

In the present section we present the $3+1$ formalism
for Einstein equations. Further details can be found in
recent books and reviews on $3+1$ numerical relativity 
\citep{gourgoulhon07, alcubierre08, baumgarte10}. 
We briefly discuss the constrained evolution schemes, focussing on
the elliptic CFC and XCFC solvers, and finally, in Sect.~\ref{sect:grmhd} we
review the GRMHD equations in $3+1$ conservative form, 
as implemented in the original ECHO scheme \citep{delzanna07}.


\subsection{The Einstein equations for conformal flatness}
\label{sect:einstein} 

The field equations of general relativity expressed in the 4D
fully covariant form
\be
G_{\mu\nu}=8\pi T_{\mu\nu},
\label{eq:einstein}
\ee
where $G_{\mu\nu}$ is the Einstein tensor containing the derivatives
of the metric tensor $g_{\mu\nu}$ and $T_{\mu\nu}$ is the matter
and/or electromagnetic energy-momentum tensor, are not
appropriate for numerical computations since time and space are
treated on an equal footing, whereas one would like to cast them in
the form of an initial value (or Cauchy) problem and evolve them
in time. 
The most widely used approach to reach this goal is based on
the so-called $3+1$ formalism, in which the generic spacetime 
$({\cal M},g_{\mu\nu})$ is split into space-like hyper-surfaces
$\Sigma_t$.  If $n^{\,\mu}$ indicates the time-like unit normal to $\Sigma_t$
(also known as the velocity of the \emph{Eulerian} observer, $n_\mu n^\mu=-1$),
the induced three-metric on each hyper-surface and the related 
\emph{extrinsic curvature} can be defined respectively as
\be
\gamma_{\mu\nu}:=g_{\mu\nu} + n_{\mu}n_{\nu},
\ee
\be
K_{\mu\nu}:= - \gamma_{\,\mu}^{\,\lambda} \der_\lambda n_\nu,
\ee
where $\der_\mu$ is the covariant derivative
with respect to $g_{\mu\nu}$ (so that $\der_\lambda g_{\mu\nu}=0$).
In general, any 4-vector or tensor can be decomposed into normal
and spatial components by contracting with $-n^{\,\mu}$ or with the
projector $\gamma^{\,\mu}_{\,\nu}$, respectively. In particular,
both $\gamma_{\mu\nu}$ and $K_{\mu\nu}$ are
purely spatial (and symmetric) tensors.

If $x^\mu := (t,x^i)$ are the spacetime coordinates adapted to 
the foliation of ${\cal M}$ introduced above, the line element 
is usually written in the so-called ADM form
\be
ds^2 := g_{\mu\nu}dx^\mu dx^\nu = -\alpha^2dt^2+
\gamma_{ij}(dx^i+\beta^idt)(dx^j+\beta^j dt),
\ee
where the \emph{lapse} function $\alpha$ and the \emph{shift vector}  $\beta^i$
(a purely spatial vector) are free \emph{gauge} functions. In this
adapted coordinate system the unit normal vector has components
$n^{\,\mu}=(1/\alpha,-\beta^i/\alpha)$ and $n_\mu=(-\alpha,0_i)$.
In the $3+1$  formalism, the Einstein equations of Eq.~(\ref{eq:einstein})
are split into a set of evolutionary equations for $\gamma_{ij}$ 
and the extrinsic curvature $K_{ij}$
\be
\partial_t \gamma_{ij} =  -2\alpha K_{ij} + \dev_i\beta_j + \dev_j\beta_i,
\label{eq:evol_gamma}
\ee
\begin{multline}
\partial_t K_{ij} = \beta^k\dev_k K_{ij} + K_{ik}\dev_j\beta^k  + K_{jk}\dev_i\beta^k  
-\dev_i\dev_j\alpha + \\ 
 \alpha [R_{ij}+KK_{ij}-2K_{ik}K_j^k] +4\pi\alpha
 [\gamma_{ij}(S-E)-2S_{ij}],
\label{eq:evol_K}
\end{multline}
plus a set of constraints that must be satisfied at all times
\be
R+K^2-K_{ij}K^{ij} = 16\pi E,
\label{eq:hamiltonian}
\ee
\be
 \dev_j (K^{ij} - K \gamma^{ij}) = 8\pi S^i,
 \label{eq:momentum}
\ee
named Hamiltonian and momentum constraints, respectively.
Here $\dev_i:=\gamma_{\,i}^{\,\mu}\der_\mu$ is the covariant derivative 
with respect to the 3-metric $\gamma_{ij}$ (so that $D_k\gamma_{ij}=0$), 
$R_{ij}$ is the Ricci tensor, again with respect to $\gamma_{ij}$, 
$R:=R_i^i$ the corresponding Ricci scalar, $K:=K_i^i$ is the trace of the 
extrinsic curvature. As far as the fluid sources are concerned
$E:=n_\mu n_\nu T^{\mu\nu}$, $S^i:= - n_{\,\mu}\gamma^i_\nu T^{\mu\nu}$,
and $S^{ij}:=\gamma^i_{\,\mu}\gamma^j_{\,\nu}T^{\mu\nu}$
(of trace $S:=S_i^i$) are, respectively, the energy density, momentum density,
and the stress-energy tensor as measured by the Eulerian observers.

The constraints introduced above are notoriously
difficult to be maintained in the numerical evolution of Eqs.~(\ref{eq:evol_gamma})
and (\ref{eq:evol_K}), and two possible approaches can be followed.
The most widely used one relies on hyperbolic formulations of the initial
value problem, the constraints are imposed only for the initial data
and numerical errors are just monitored or damped during
time evolution (\emph{free evolution schemes}).
On the other hand, the constraints can be enforced at each timestep during
the numerical simulation, leading to the so-called
\emph{constrained evolution schemes}, where the main idea is to maximize
the number of elliptic equations, usually more stable than hyperbolic
equations. Moreover, in the steady state case the set of equations should
easily reduce to those used for stationary spacetimes and for the construction of
initial data. An example is the so-called \emph{fully constrained formalism} (FCF)
for asymptotically flat spacetimes in full GR \citep{bonazzola04}, which contains
the widely used set of CFC elliptic equations as an approximation.
In the following we will describe the general assumptions of conformal flatness 
and we will review the set of CFC equations.

Let us start by applying a Lichnerowicz conformal decomposition
\be
\gamma_{ij}:=\psi^4 f_{ij},
\label{eq:f}
\ee
assuming a flat background metric $f_{ij}$ (\emph{time independent} and not necessarily
in Cartesian coordinates), where the conformal factor 
satisfies $\psi=(\gamma/f)^{1/12}$, with $f:={\rm det} f_{ij}$.
A second assumption is the condition of \emph{maximum slicing} of foliations
\be
K=0.
\label{eq:slicing}
\ee
Under these assumptions, to be preserved during time evolution,
the trace of Eq.~(\ref{eq:evol_gamma}) and its traceless part become, respectively
\be
\partial_t\ln\gamma^{1/2} \equiv \partial_t\ln\psi^{1/6}=D_i\beta^i,
\label{eq:evol_psi}
\ee
\be
2\alpha K_{ij}=
D_i\beta_j + D_j\beta_i -\textstyle{\frac{2}{3}}(D_k\beta^k)\gamma_{ij},
\label{eq:K}
\ee
where $\partial_t\gamma=0$ if, and only if, $D_i\beta^i=0$
and the extrinsic curvature can be expressed in terms 
of derivatives of the shift vector alone.
The next step is to use covariant derivatives associated to
the flat 3-metric $f_{ij}$, that will be indicated here by the
usual \emph{nabla} operator $\der_i$ (and $\der_k f_{ij}=0$). 
When Eq.~(\ref{eq:f}) holds, it is possible to demonstrate
that the Ricci scalar for $\gamma_{ij}$ (that for $f_{ij}$ is zero) is
\be
R=-8\psi^{-5}\Delta\psi,
\ee 
in which $\Delta:=\der_i\der^i$ is the usual Laplacian of flat space.
The above relation combined first with the Hamiltonian constraint
in Eq.~(\ref{eq:hamiltonian}) and then with the trace of Eq.~(\ref{eq:evol_K}) 
provides the following two scalar Poisson-like equations
for the conformal factor $\psi$ and the lapse function $\alpha$
\be
\Delta \psi = - [ 2\pi E+\textstyle{\frac{1}{8}} K_{ij}K^{ij} ] \,\psi^5,
\ee
\be
\Delta (\alpha\psi) = [ 2\pi (E+2S) + \textstyle{\frac{7}{8}} K_{ij}K^{ij} ] \,\alpha\psi^5,
\ee
where we still need to write $K_{ij}$ in terms of flat space derivatives
of the shift vector.

To the traceless extrinsic curvature 
$A_{ij}:=K_{ij}-\textstyle{\frac{1}{3}}K\gamma_{ij}\equiv K_{ij}$ 
is then applied a conformal \emph{time-evolution} rescaling
\be
K^{ij}=\psi^{-4} \Atil^{ij}, ~~~ 2\alpha\Atil^{ij} := (L\,\beta)^{ij},
\label{eq:cts}
\ee
where the \emph{conformal Killing} operator associated to the flat metric and
applied to the vector $\beta^i$ is defined as
\be
(L\,\beta)^{ij} := \der^i\beta^j + \der^j\beta^i - \textstyle{\frac{2}{3}} (\der_k \beta^k)f^{ij}.
\ee
This scaling is also employed in the so-called 
\emph{conformal thin sandwich} (CTS)  approach to initial data.
Moreover, it is quite natural, due to Eq.~(\ref{eq:K}), and 
since within a conformally flat decomposition of the metric we have
\be
(L_\gamma \beta)^{ij}=\psi^{-4} (L \,\beta)^{ij},
\label{eq:L}
\ee 
where $L_\gamma$ indicates here the conformal Killing operator 
associated to $\gamma_{ij}$.
On the other hand, in conformal flatness we also have
\be
D_j K^{ij}=\psi^{-10}\der_j (\psi^{10}K^{ij}),
\label{eq:confmom}
\ee
to be used with the above rescaling in the momentum constraint
to find an equation for $\beta^i$.

Thanks to all the relations derived so far, the final set of CFC elliptic
equations may be written in terms of the sources and of $\Atil^{ij}$ 
(containing $\alpha$ and first derivatives of $\beta^i$) as
\be
\Delta \psi = - [ 2\pi E+
\textstyle{\frac{1}{8}}f_{ik}f_{jl}\Atil^{ij}\Atil^{kl} ] \,\psi^5 ,
\label{eq:cfc_psi}
\ee
\be
\Delta (\alpha\psi) = [ 2\pi (E+2S) +
 \textstyle{\frac{7}{8}} f_{ik}f_{jl}\Atil^{ij}\Atil^{kl} ] \,\alpha\psi^5,
\label{eq:cfc_alpha}
\ee
\be
\Delta_L \,\beta^i = 16\pi \alpha \psi^4 S^i + 2\psi^6 \Atil^{ij}\der_j (\alpha\psi^{-6}),
\label{eq:cfc_beta}
\ee
where
\be
\Delta_L \,\beta^i:=\der_j (L\,\beta)^{ij}=
\Delta \beta^i +\textstyle{\frac{1}{3}}\der^i (\der_j \beta^j),
\ee
is the so-called \emph{conformal vector Laplacian} operator,
associated to the flat 3-metric $f_{ij}$ and applied to $\beta^i$.


\subsection{From CFC to XCFC}
\label{sect:xcfc}

A slightly different approach to the Einstein equations for 
asymptotically flat spacetimes has been recently 
presented \citep{cordero-carrion09}. This involves a rewriting
of the elliptical part of the FCF system for full GR through a different
decomposition of the extrinsic curvature. 
Here we will just describe its conformal flatness approximation,
leading to the so-called \emph{extended conformal flatness condition}
(XCFC) system of elliptic equations, improving on the CFC ones
described in the previous section. The set of XCFC equations
is our choice for the metric evolution in X-ECHO.

The new approach still relies on the usual conformal decomposition
in Eq.~(\ref{eq:f}) and on the maximum slicing condition of Eq.~(\ref{eq:slicing}), 
but the choice for the decomposition of the (traceless) extrinsic curvature is different.
We use here the \emph{momentum-constraint} rescaling and the so-called
York \emph{conformal transverse traceless}  (CTT) decomposition, first
introduced for initial data, that is
\be
K^{ij} = \psi^{-10} \Ahat^{ij}, ~~~ \Ahat^{ij}  := (LW)^{ij} +\Ahat^{ij}_{TT},
\label{eq:ctt}
\ee
where the conformal Killing operator associated to the unknown vector
$W^i$  gives the \emph{longitudinal} part of $\Ahat_{TT}^{ij}$, whereas
 $\Ahat_{TT}^{ij}$ is a transverse ($\der_j\Ahat_{TT}^{ij}=0$),
traceless ($f_{ij}\Ahat_{TT}^{ij}=0$) tensor. 
Consistency between the CTS and CTT decompositions 
(notice that $\Ahat^{ij}=\psi^6\Atil^{ij}$) should require a 
non-vanishing $\Ahat_{TT}^{ij}$. However it has been demonstrated that
this quantity is even smaller than the non-conformal part of the
spatial metric within the CFC approach, and hence can be safely
neglected on the level of the CFC approximation.
Thus we set, as an additional hypothesis
\be
\Ahat_{TT}^{ij}=0 \Rightarrow \Ahat^{ij} = (L W)^{ij},
\ee
so that $\Ahat^{ij}$ is defined in terms of the auxiliary vector $W^i$ alone.
The latter is to be derived from the momentum constraint 
using Eq.~(\ref{eq:confmom}), that is simply
\be
\der_j \Ahat^{ij} = \Delta_L W^i =8\pi\,\psi^{10}S^i,
\ee
to be added to the other CFC equations.

The final \emph{augmented} set of CFC elliptic equations, also
known as XCFC equations, is then the following
\be
\Delta_L W^i = 8\pi f^{ij}\Sbar_j,
\ee
\be
\Delta \psi = -  2\pi\Ebar\, \psi^{-1}
- \textstyle{\frac{1}{8}}f_{ik}f_{jl}\Ahat^{kl}\Ahat^{ij}\, \psi^{-7} ,
\label{eq:xcfc_psi}
\ee
\be
\Delta (\alpha\psi) = [2\pi (\Ebar+2\Sbar)\, \psi^{-2} +
\textstyle{\frac{7}{8}} f_{ik}f_{jl}\Ahat^{kl}\Ahat^{ij} \, \psi^{-8} ] \, \alpha\psi,
\label{eq:xcfc_alpha}
\ee
\be
\Delta_L \,\beta^i = 16\pi \,\alpha\psi^{-6} f^{ij}\Sbar_j 
+ 2\Ahat^{ij}\der_j (\alpha\psi^{-6}),
\label{eq:xcxf_beta}
\ee
where for convenience we have introduced rescaled fluid 
source terms of the form
\be
\Sbar_j:=\psi^6 S_j,\,\,\, \Ebar:=\psi^6 E,\,\,\, \Sbar:=\psi^6 S,
\ee
and we remind that
\be
\Ahat^{ij} =\der^iW^j + \der^jW^i - \textstyle{\frac{2}{3}} (\der_k W^k)f^{ij}.
\label{eq:Ahat}
\ee
Some comments and comparisons between the CFC and XCFC
sets of equations are now due.
\begin{itemize}
\item The unknown functions are now 8 rather than 5 ($W^i$, $\psi$,
$\alpha$, $\beta^i$), and this is reflected by the augmented number
of elliptic equations (there is a new vector Poisson equation for the
auxiliary variable $W^i$).
\item While in CFC all the equations were strongly coupled, here
the equations can be solved hierarchically one by one, in the
given order, since each right hand side just contains
known functions or the variable itself (in the two scalar 
Poisson-like equations for $\psi$ and $\alpha\psi$).
\item As we will see in the next sub-section, schemes for general
relativistic hydrodynamics or MHD (like ECHO), given
a metric in $3+1$ form, actually evolve in time the
\emph{conservative} variables $\gamma^{1/2}S_j$ and 
$\gamma^{1/2}E$, rather than $S^i$ and $E$.
Since $\psi^6=\gamma^{1/2}/f^{1/2}$ and $f^{1/2}$ is known
and time-independent, the sources $\Sbar_j$ and $\Ebar$
are basically known after each computational time-step
without the need of an updated value of $\psi$. This will
be only needed to work out $\Sbar=\psi^6\gamma_{ij}S^{ij}$,
after the new value of $\psi$ has been provided by
Eq.~(\ref{eq:xcfc_psi}) and the inversion of conservative 
to \emph{primitive} variables has been achieved.
Primitive variables are then updated self-consistently
together with the new values for the metric, whereas this
was not possible in CFC. In that case, one could either use
Eq.~(\ref{eq:evol_psi}) to derive a guess of the updated
$\psi$ (a method easily prone to both convergence problems
and discretization errors), or one is forced to iterate simultaneously 
over the metric solver (the whole CFC set) and the inversion
routine for the primitive variables (typically itself a numerical
iterative Raphson-Newton method).
\item The last, and certainly not least, issue is related to the
mathematical nature of the scalar Poisson-like equations.
In both cases we have a structure of the form
\be
\Delta u = h u^p,
\label{eq:poiss}
\ee
where $u$ is the generic variable ($\psi$ or $\alpha\psi$),
$h$ is the generic source term, and $p$ provides the exponent of the 
non-linearity ($p=0$ for a canonical Poisson equation). It can be demonstrated
that the condition $ph \geq 0$ implies that the solution
$u$ is \emph{locally unique}.  While this is always true
in XCFC, since we have two contributions with $p=-1$ and $p=-7$,
both with $h\leq 0$, in Eq.~(\ref{eq:xcfc_psi}), and one
contribution with $p=+1$ and $h\geq 0$ in Eq.~(\ref{eq:xcfc_alpha}),
local uniqueness cannot be guaranteed for the CFC system,
since Eq.~(\ref{eq:cfc_alpha}) contains a term that certainly
violates the requirement (the second one, due to the presence
of a factor $\alpha^{-1}$ in $\Atil^{ij}$).
\end{itemize} 


\subsection{The GRMHD equations and the ECHO scheme}
\label{sect:grmhd}

The equations for an ideal, magnetized, perfectly conducting
plasma are
\be
\nabla_{\mu} (\rho u^{\,\mu})=0,
\ee
\be
\nabla_{\mu}T^{\mu\nu}=0,
\ee
\be
\nabla_{\mu}F^{*\mu\nu} = 0, 
\ee
respectively the continuity equation, the conservation law for
momentum-energy, and the sourceless Maxwell equation.
Here $\rho$ is the mass density as measured in the frame comoving 
with the fluid 4-velocity $u^{\,\mu}$, the total momentum-energy tensor is
\be
T^{\mu\nu}=\rho h\,u^{\,\mu}u^{\nu}+pg^{\,\mu\nu} +
{F^{\mu}}_{\lambda}F^{\nu\lambda}-
\textstyle{\frac{1}{4}}(F^{\lambda\kappa}F_{\lambda\kappa})g^{\,\mu\nu},
\label{eq:Ttot}
\ee
with $h=1+\varepsilon + p/\rho$ the specific enthalpy, $\epsilon$ the specific internal 
energy, $p=p(\rho,\varepsilon)$ the thermal pressure (provided by some form of
equation of state, EoS). Moreover, $F^{\mu\nu}$ is the Faraday (antisymmetric) 
electromagnetic tensor, with the associated dual 
$F^{*\mu\nu}=\frac{1}{2}\epsilon^{\,\mu\nu\lambda\kappa}F_{\lambda\kappa}$,
where $\epsilon^{\,\mu\nu\lambda\kappa}=(-g)^{-1/2}[\mu\nu\lambda\kappa]$ 
is the spacetime Levi-Civita pseudo-tensor 
($\epsilon_{\mu\nu\lambda\kappa}=-(-g)^{1/2}[\mu\nu\lambda\kappa]$), 
with $g=\mathrm{det} g_{\mu\nu}$ and $[\mu\nu\lambda\kappa]$ 
is the alternating Levi-Civita symbol.
The system is closed by Ohm's law for a perfectly conducting plasma,
which becomes a constraint for a vanishing electric field in the frame 
comoving with the fluid
\be
F^{\mu\nu}u_{\nu}=0.
\label{eq:mhd}
\ee
This basically replaces the Maxwell equation $\nabla_{\mu}F^{\mu\nu} = -J^\nu$,
where the 4-current $J^\nu$ is a derived quantity as in classical MHD.

In order to derive the GRMHD equations in $3+1$ form, as employed in ECHO,
we must decompose all 4-vectors and tensors into their spatial 
and temporal components on each slice $\Sigma_t$ of the time evolution.
This can be easily achieved by using the unit normal vector $n^{\,\mu}$
introduced in the previous section
\be
\rho  u^{\,\mu}  :=  D ( v^{\,\mu} + n^{\,\mu} ) , 
\label{eq:u}
\ee
\be
T^{\mu\nu}  :=  S^{\mu\nu} + n^{\,\mu}S^{\nu} + S^{\mu}n^{\nu} + En^{\,\mu}n^{\nu},  
\label{eq:T}
\ee
\be
F^{\mu\nu}  :=  n^{\,\mu}E^{\nu} - E^{\mu}n^{\nu} + 
\epsilon^{\,\mu\nu\lambda\kappa}B_{\lambda}n_{\kappa}, 
\label{eq:F}
\ee
\be
F^{*\mu\nu}  :=  n^{\,\mu}B^{\nu} - B^{\mu}n^{\nu}  - 
\epsilon^{\,\mu\nu\lambda\kappa}E_{\lambda}n_{\kappa}, 
\label{eq:F*}
\ee
where every new quantity is now purely spatial, as measured by the Eulerian observer.
In particular, $D:=\rho\Gamma$ is the rest mass density, $v^i$ is the fluid velocity, 
$\Gamma:=(1-v_iv^i)^{-1/2}$ is the usual Lorentz factor, whose definition is due to the
condition $u_\mu u^{\,\mu}=-1$. The stress-energy 3-tensor $S^{ij}$, its trace $S$, the
momentum density $S^i$, and the energy density $E$ are the same 
quantities appearing in the $3+1$ Einstein equations (this is why notations have been 
slightly modified with respect to the original ECHO paper), and are respectively given by
\be
S_{ij}=\rho h \Gamma^2 v_iv_j + p\gamma_{ij} - E_iE_j - B_iB_j + 
\textstyle{\frac{1}{2}}(E_kE^k+B_kB^k)\gamma_{ij},
\ee
\be
\hspace{-2mm}S=\rho h (\Gamma^2-1) + 3p + \textstyle{\frac{1}{2}}(E_iE^i+B_iB^i),
\ee
\be
\hspace{-2mm}S_i=\rho h \Gamma^2 v_i + \epsilon_{ijk}E^jB^k,
\ee
\be
\hspace{-2mm}E=\rho h \Gamma^2 - p + \textstyle{\frac{1}{2}}(E_iE^i+B_iB^i).
\ee
The electric and magnetic fields as measured by the Eulerian observer
are defined as $E^i:= - n_{\,\mu} \gamma^i_\nu F^{\mu\nu}$ and 
$B^i:= - n_{\mu}\gamma^i_\nu F^{*\mu\nu}$, and due to the 
condition in Eq.~(\ref{eq:mhd}), the electric field is a derived 
quantity precisely as in classical MHD
\be
E_i=-\epsilon_{ijk}v^jB^k,
\ee
where $\epsilon_{ijk}=\gamma^{1/2}[ijk]$ ($\epsilon^{ijk}=\gamma^{-1/2}[ijk]$) 
is the Levi-Civita pseudo-tensor for the 3-metric $\gamma_{ij}$, 
and $[ijk]$ is the alternating symbol taking values $+1$, $-1$, or $0$.

Thanks to the above decompositions, the GRMHD equations can be 
entirely rewritten in terms of purely spatial vectors, while retaining 
the original conservation form.
We end up with a sub-set of fluid-like balance laws in
divergence form
\be
\partial_t \vec{\mathcal{U}} + \partial_i \vec{\mathcal{F}}^i = \vec{\mathcal{S}},
\label{eq:dudt}
\ee
plus a magnetic sub-set with the induction equation in curl form
and the associated divergence-free condition, to be preserved
at all times during evolution
\be
\partial_t \mathcal{B}^i + [ijk] \partial_j \mathcal{E}_k = 0,~~~\partial_i \mathcal{B}^i=0.
\label{eq:dbdt}
\ee
The set of \emph{conservative} fluid variables and the set of associated fluxes
are, respectively
\be
\vec{\mathcal{U}} := \gamma^{1/2}\!\left[\begin{array}{c}
\!D\! \\ \!S_j\! \\ \!E\!
\end{array}\right],~~~
\vec{\mathcal{F}}^i := \gamma^{1/2}\!\left[\begin{array}{c}
\! (\alpha v^i - \beta^i) D \! \\
\!\alpha S^i_j-\beta^i S_j\! \\
\!\alpha S^i-\beta^i E\!
\end{array}\right],
\ee
whereas the set of source terms contain the derivative of the metric
and thus the curvature effects
\be
\vec{\mathcal{S}} := \gamma^{1/2}\!\left[\begin{array}{c}
0 \\  
\frac{1}{2}\alpha S^{ik}\partial_j\gamma_{ik}+
S_i\partial_j\beta^i-E\partial_j\alpha \\ 
\alpha K_{ij}S^{ij}-S^j\partial_j\alpha\!
\end{array}\right]\!.
\ee
Here $K_{ij}$ is the extrinsic curvature introduced in the previous
section, whose evolution is directly provided by the Einstein 
equations in the $3+1$ formalism, together with $\gamma_{ij}$,
or may be given in terms of the derivatives of the metric terms.
For a dynamical spacetime under the maximal slicing condition,
from Eq.~(\ref{eq:K}) we can write
\be
\alpha K_{ij}S^{ij}=\textstyle{\frac{1}{2}}\beta^jS^{ik}\partial_j\gamma_{ik} + 
S_i^j\partial_j\beta^i - \textstyle{\frac{1}{3}} [\gamma^{-1/2}\partial_i(\gamma^{1/2}\beta^i)] S,
\ee
and the same expression without the last term 
may be used for any stationary spacetime (Cowling approximation).
As far as the induction equation is concerned, we have defined
\be
\mathcal{B}^i \! := \! \gamma^{1/2}B^i,
\ee
\be
\mathcal{E}_i \! := \! \alpha E_i + \epsilon_{ijk}\beta^jB^k
\! = \! - \gamma^{1/2}[ijk](\alpha v^j-\beta^j)B^k,
\ee
where the vector $\alpha v^i-\beta^i$ is sometimes called transport velocity.
The induction equation may be also written in the equivalent
divergence form as
\be
\partial_t \mathcal{B}^j + \partial_i 
\{ \gamma^{1/2} [ (\alpha v^i - \beta^i)B^j - B^i(\alpha v^j - \beta^j)] \}=0,
\ee
and the antisymmetric nature of the magnetic fluxes reflects
that of the original electromagnetic tensor.
Whatever is the adopted choice for the form of the induction
equation, we have a final set of eight hyperbolic equations.
Usually the corresponding variables are named \emph{primitive}
variables, for example the set
\be
\vec{\mathcal{P}}:=[\rho,v^i,p,B^i]^T,
\ee
for which $\vec{\mathcal{U}}=\vec{\mathcal{U}}(\vec{\mathcal{P}})$, 
$\vec{\mathcal{F}}^i=\vec{\mathcal{F}}^i(\vec{\mathcal{P}})$, and
$\vec{\mathcal{S}}=\vec{\mathcal{S}}(\vec{\mathcal{P}})$, where
for simplicity we consider here the augmented system with $\mathcal{B}^j$
in $\vec{\mathcal{U}}$.

As discussed in the previous section, the inversion of the
non-linear system
$\vec{\mathcal{U}}=\vec{\mathcal{U}}(\vec{\mathcal{P}})$
to recover the set of primitive variables is achieved through
a numerical iterative scheme and requires the knowledge
of the volume element $\gamma^{1/2}$, for consistency
updated at the same time level as $\vec{\mathcal{U}}$.
Moreover, as discussed in \citet{delzanna07}, the conservative to 
primitive variables inversion is the most delicate part of a relativistic MHD code,
as high Lorentz factor flows or strong magnetic fields (i.e. in a NS
magnetosphere) may easily lead to errors in the values of the
conservative variables. The whole procedure can
be reduced to the solution of two coupled nonlinear equations,
for any given EoS.
In X-ECHO we leave complete freedom in the choice
$p=p(\rho,\varepsilon)$, however in the present paper, given
the nature of the numerical tests proposed, we limit this choice
to either an ideal $\gamma$-law EoS
\be
p(\rho,\varepsilon)=(\gamma-1)\rho\varepsilon,
\ee
or  to a polytropic law EoS
\be
p(\rho)=K\rho^\gamma,\quad,
\varepsilon (\rho)=\frac{1}{\gamma-1}K\rho^{\gamma-1},
\ee
where $\gamma$ and $K$ are given constants.
When the first option is used, the root-finding procedure
solves for the variable $x=v_iv^i$, according to the third
method described in the ECHO paper, where the nested
second equation (a cubic) for $y=\rho h\Gamma^2$ can be
either solved analytically or iteratively with an inner loop
(and we typically adopt this latter choice). Notice that when
$S_iB^i=0$, or in general for a purely fluid simulation,
the equation for $y$ is linear and can be readily solved.
On the other hand, for a polytropic law the energy equation
becomes redundant, since all thermodynamical quantities
are now functions of the density alone, and the overall
inversion procedure simply reduces to a root-finding iterative
solver for $x$.


\section{The XCFC solver for axisymmetric spacetimes}
\label{sect:x-echo}

The X-ECHO code for GRMHD in dynamical spacetimes
is built upon the ECHO scheme, coupled to a novel solver
for the XCFC equations described in Sect.~\ref{sect:xcfc}.
All the necessary definitions and equations have been provided 
in the previous section, here we specialize to the particular
implementation of X-ECHO we are mostly interested in,
that is under the assumption of axisymmetric GRMHD
configurations, adopting spherical-like coordinates 
$x^i = (r,\theta,\phi)$ for the conformal flat metric.

Thus, as a first step we assume in Eq.~(\ref{eq:f}) the usual
spherical coordinates
\be
f_{ij}=\mathrm{diag}(1,r^2,r^2\sin^2\theta),
\label{eq:f_sph}
\ee
so that
\be
ds^2=-\alpha^2 dt^2+\psi^4[(dr+\beta^rdt)^2+
r^2(d\theta+\beta^\theta dt)^2+r^2\sin^2\theta(d\phi+\beta^\phi dt)^2],
\label{eq:cfc_sph}
\ee
with $f^{1/2}=r^2\sin\theta$ and $\gamma^{1/2}=\psi^6\,r^2\sin\theta$.
Moreover, we are going to specialize to the axisymmetric case, thus
the condition
\be
\partial_\phi \equiv 0,
\ee
will be assumed throughout the paper.
Within the flat metric $f_{ij}$, it is convenient to introduce the orthonormal basis 
$\vec{e}_{\hat i}:=(\vec{e}_{\hat r},\vec{e}_{\hat\theta},\vec{e}_{\hat\phi})$, with
\be
\vec{e}_{\hat r}:=\partial_r,~~~~
\vec{e}_{\hat\theta}:=r^{-1}\partial_\theta,~~~~
\vec{e}_{\hat\phi}:=(r\sin\theta)^{-1}\partial_\phi,
\ee
for which $f_{{\hat i}{\hat j}}=\mathrm{diag}(1,1,1)$, where a similar
notation as in \citet{bonazzola04} has been assumed.
Any generic vector $\vec{X}$ can be then expressed in the usual form as
\be
\vec{X}:=X^{\hat r} \vec{e}_{\hat r} + X^{\hat\theta} \vec{e}_{\hat\theta} 
+ X^{\hat\phi} \vec{e}_{\hat\phi},
\ee
where the orthonormal (with respect to $f_{ij}$) components $X^{\hat i}$ are, 
respectively
\be
X^{\hat r}:=X^r,~~~X^{\hat \theta} := r\, X^\theta,~~~X^{\hat \phi} := r\sin\theta\, X^\phi,
\ee
while the relation to covariant components still involves the function
$\psi$, since $X_i=\psi^4 f_{ij}X^j$.
As far as covariant derivatives are concerned, the change of basis allows
one to use the $\nabla_{\hat i}$ operator of spherical coordinates.
In particular, the Laplacian of a generic scalar function $u(r,\theta)$ is
\be
\Delta u = \partial_r^2 u +2r^{-1} \partial_r u +
r^{-2}(\partial_\theta^2 u +\cot\theta\, \partial_\theta u),
\label{eq:m1}
\ee
whereas the orthonormal components of the conformal vector
Laplacian are, respectively
\be
(\Delta_L \vec{X})^{\hat r} = \Delta X^{\hat r} - 2r^{-2} (X^{\hat r} + 
\partial_\theta  X^{\hat\theta}
+ \cot\theta\, X^{\hat\theta} ) + \textstyle{\frac{1}{3}}\partial_r (\nabla \cdot \vec{X}),
\label{eq:m2}
\ee
\be
(\Delta_L \vec{X})^{\hat\theta} = \Delta X^{\hat\theta} + 2r^{-2} \partial_\theta  X^{\hat r}
- (r\sin\theta)^{-2}X^{\hat\theta} + \textstyle{\frac{1}{3}}r^{-1}\partial_\theta (\nabla \cdot \vec{X}),
\label{eq:m3}
\ee
\be
\hspace{-5mm} (\Delta_L \vec{X})^{\hat\phi} = \Delta X^{\hat\phi} - (r\sin\theta)^{-2}X^{\hat\phi},
\label{eq:m4} 
\ee
where the divergence of $\vec{X}$ is
\be
\nabla \cdot \vec{X} = \partial_r X^{\hat r} + 
r^{-1} (2 X^{\hat r} + \partial_\theta X^{\hat\theta} + \cot\theta\, X^{\hat\theta}),
\label{eq:m5}
\ee
precisely the formulae of vector calculus in spherical coordinates.

The Poisson-like elliptic equations used in XCFC to compute the
eight metric terms consist of two of scalar equations in the
form of Eq.~(\ref{eq:poiss}), for the variables $u=\psi$ and
$u=\alpha\psi$, and two vector equations for the generic
unknown vector $X^i=W^i$ and $X^i=\beta^i$.
Due to non-linearity, the scalar equations are better solved
iteratively for the quantity $q_n:=u_n-1$, which in both cases gives
the deviation from asymptotic flatness $\psi\to 1$ and $\alpha\to 1$,
where the new value at step $n$ is computed using the previous value at
step $n-1$ in the source term, until convergence is reached within
some prescribed tolerance.
Summarizing, the metric equations are expressed in one
of the two generic forms
\be
\Delta \, q_n = H_{n-1}\equiv h\, (1+q_{n-1})^p,
\label{eq:poiss_it}
\ee
and, using the orthonormal basis introduced above,
\be
(\Delta_L \vec{X})^{\hat i} = H^{\hat i},
\label{eq:vec_poiss}
\ee
where $h$ and $\vec{H}$ are generic scalar and vector source terms,
to be provided by the XCFC equations, so that both the right hand
sides above are known functions of $(r,\theta)$.

Numerical methods available for solving these elliptic
partial differential equations (PDEs) can be divided into three main 
categories \citep[see, for methods and discussions][]{grandclement01, 
dimmelmeier05, grandclement09}: direct inversion, full relaxation, spectral. 
Direct inversion codes are able to solve the complete system
of CFC (or XCFC) at once using Newton-Raphson solver (or any other
inversion technique) on the entire computational grid over which the
metric solution is desired. They have very good convergence to machine 
accuracy within few steps, but they suffer serious limitations: 
the initial guess must be close enough to the solution to avoid convergence
on local minima (instead of global minima); memory requirement for
matrix allocation is typically very large (usually a sparse matrix is needed
in the whole 2D domain); in general direct
inversion schemes solve the metric on smaller grids that the one over
which the fluid variables are evolved, and require interpolation between the
two, with problems that may arise at boundaries.
Full relaxation codes use SOR, multigrid or other relaxation techniques. The
schemes are fast, they require little memory allocation, but usually
suffer from poor convergence properties: the convergence is in
general slow (compared to direct or spectral schemes) and
might fail on the axis or at the center, due to the singular nature of
some metric elements.
Spectral schemes decompose the set of CFC (or XCFC) equations using a
combination of spherical harmonics (based on Legendre polynomials) 
in the angular directions and Chebyshev polynomials in the radial direction.
This ensures a correct behavior on the axis and at the center even with
a limited number of eigenfunctions, but they require specific grids
and sometimes complex compactifications or multi-domain decomposition
techniques with appropriate boundary conditions for each multipole 
and each domain.
The metric solver in X-ECHO uses a mixed technique. In the angular direction we
decompose using spherical harmonics, to preserve the correct
asymptotic form on axis. However, the set of ordinary differential equations
(ODEs) obtained for each harmonic is then solved using direct 
inversion over the same radial grid used in the GRMHD code, with no need of 
interpolation or compactification. At second order accuracy in a finite difference
discretization, the scalar equations reduce to the simple inversion of tridiagonal
matrices (band diagonal matrices for the poloidal components of the
vector Poisson equations), where appropriate solvers are fast, require 
little memory allocation, and typically converge with high accuracy.

For the scalar Poisson-like Eq.~(\ref{eq:poiss_it}) we then decompose,
for each level $n$ of iteration (here omitted), the unknown $q$ as
\be
q(r,\theta):=\sum_{l=0}^\infty \left[ A_l(r) Y_l(\theta) \right],
\ee
where 
\be
Y_l(\theta)\equiv Y_l^0(\theta):=\textstyle{\sqrt{\frac{2l+1}{4\pi}}}P_l(\cos\theta),
\ee
with $P_l$ the Legendre polynomial of degree $l$ and the axisymmetry 
condition has been imposed ($m=0$).
The PDE, where the Laplacian is provided in Eq.~(\ref{eq:m1}), 
may be then split into the series of radial ODEs for each harmonic $l$
\be
\frac{d^2 A_l}{dr^2} + \frac{2}{r}\frac{d A_l}{dr} -\frac{l(l+1)}{r^2}A_l  =H_l,
\ee
where the new source term is
\be
H_l(r):=\oint H(r,\theta) Y_l(\theta) d\Omega,
\label{eq:int}
\ee
with $d\Omega=2\pi \sin\theta d \theta$ and the integral running from
$\theta=0$ to $\theta=\pi$ due to axisymmetry.

As far as the vector Poisson equation in Eq.~(\ref{eq:vec_poiss})
is concerned, the unknown vector $\vec{X}$ is first decomposed
into vector spherical harmonics, that is, in the axisymmetric case
\be
\vec{X}(r,\theta):= \sum_{l=0}^\infty \left[ A_l(r) Y_l(\theta)\vec{e}_{\hat r} + 
B_l(r) Y^\prime_l(\theta)\vec{e}_{\hat\theta} +
C_l(r) Y^\prime_l(\theta)\vec{e}_{\hat\phi} \right],
\ee
where $Y^\prime_l := dY_l/d\theta$. As it is apparent from
the operators in Eqs.~(\ref{eq:m2}-\ref{eq:m5}), the set of
equations split into a series of ODEs with, for each harmonic $l$, 
a coupled poloidal part for the radial functions $A_l(r)$ and $B_l(r)$ 
\be
\!\frac{4}{3}\frac{d^2 A_l}{dr^2} + \frac{8}{3r^2}\!\left(r\frac{d A_l}{dr} - A_l \right)
- \frac{l(l \!+\! 1)}{r^2}\! \left(A_l + \frac{r}{3}\frac{dB_l}{dr}  - \frac{7}{3}B_l \right)=H_l^{\hat r},
\ee
\be
\frac{d^2 B_l}{dr^2} + \frac{2}{r}\frac{d B_l}{dr} -\frac{4l(l \!+\! 1)}{3r^2}B_l 
+ \frac{1}{3r}\frac{dA_l}{dr} + \frac{8}{3r^2}A_l =H_l^{\hat\theta},
\ee
and a toroidal part
\be
\frac{d^2 C_l}{dr^2} + \frac{2}{r}\frac{d C_l}{dr} -\frac{l(l+1)}{r^2}C_l  =H_l^{\hat\phi},
\ee
for $C_l(r)$ alone.
The new source terms are given by
\be
H_l^{\hat r}(r):=\oint H^{\hat r}(r,\theta) Y_l(\theta) d\Omega,
\ee
\be
H_l^{\hat\theta}(r):=\frac{1}{l(l+1)}\oint H^{\hat\theta}(r,\theta) Y^\prime_l(\theta) d\Omega,
\ee
\be
H_l^{\hat\phi}(r):=\frac{1}{l(l+1)}\oint H^{\hat\phi}(r,\theta) Y^\prime_l(\theta) d\Omega.
\ee
These integrals, as well as that in Eq.~(\ref{eq:int}), are computed in X-ECHO 
by using Gaussian quadrature points, so the original source terms must be 
first interpolated on the required locations.

As anticipated, we use finite differences and a second order approximation
for first and second spatial derivatives, so that, for each harmonic $l$,
the two coupled poloidal equations are reduced to the inversion of a sparse 
matrix of bandwidth 7, whereas the toroidal equation leads to a tridiagonal 
matrix (standard open source routines are employed).
Boundary conditions at $r = 0$ and $r = r_\mathrm{max}$ 
are given by the parity and asymptotic properties of the multipole
corresponding to the harmonic $l$. In particular, we assume that at
the center $A_l(r)$ has parity $(-1)^l$, whereas $B_l(r)$ and $C_l(r)$
have parity $(-1)^{l+1}$, and at large distances from the central sources 
the multipoles  are forced to decay as $r^{-l(l+1)}$.


\section{Axisymmetric GRMHD equilibria: the XNS code}
\label{sect:xns}

One of the most common application of the fully constrained formalism 
is the search of self-consistent (axisymmetric) equilibrium configurations 
(i.e. fluid quantities and metric) for compact relativistic stars (we will simply
refer to these objects with the term NS) a typical case of initial data problem 
in GR \citep{cook00, stergioulas03, gourgoulhon10}.
Several codes have been presented in the years addressing this issue 
\citep{komatsu89a, komatsu89b, cook94,  stergioulas95, 
nozawa98, bonazzola98, kiuchi08a},
and all of them, despite the different approaches and upgrades 
(e.g. differential rotation, toroidal magnetic field), adopt the so-called 
{\it quasi-isotropic} coordinates. Under the conditions
\be
\partial_t\equiv 0, \quad \partial_\phi\equiv 0 ,
\ee
we write here the corresponding line element in the form
\be
ds^2 = -\alpha^2dt^2+\psi^4 (dr^2+r^2d\theta^2) + R^2(d\phi -\omega dt)^2,
\label{eq:qisotropic}
\ee
which resembles that of the CFC metric for axisymmetric spacetimes 
of Eq.~(\ref{eq:cfc_sph}) in spherical coordinates and reduces to it when
\be
R=\psi^2r\sin\theta,
\label{eq:R}
\ee
where only in this case the function $\psi$ in Eq.~(\ref{eq:qisotropic})
recover the meaning of conformal factor. In general, we can think
the metric function $R:=\gamma_{\phi\phi}^{1/2}$ to be a sort of generalized 
cylindrical radius, whereas $\omega:=-\beta^\phi$ is the intrinsic
angular velocity about the symmetry axis of the 
\emph{zero angular momentum observers} \citep[ZAMOs:][]{bardeen72},
that are the normal (Eulerian) observers for an axisymmetric spacetime.
When $\omega=0$ the spacetime is spherically symmetric and the
two metrics both reduce to that in isotropic Schwarzschild coordinates.
In the following we briefly re-derive the GRMHD equilibrium condition 
(a Bernoulli-like integral) for purely toroidal flows and magnetic fields in
quasi-isotropic coordinates, more general derivations can be
found in \citep{kiuchi08a} or \citep{komissarov06b}, this latter
work applied to magnetized tori around rotating BHs
\citep[see also the final numerical test in][]{delzanna07}. 

The only non vanishing components of the spatial (Eulerian) 3-vectors 
$v^i$ and $B^i$ are the azimuthal ones, and the corresponding moduli are
\be
v:=(v_\phi v^\phi)^{1/2}=Rv^\phi=\alpha^{-1}R(\Omega-\omega),
\label{eq:v}
\ee
\be
B:=(B_\phi B^\phi)^{1/2}=RB^\phi,
\ee  
where we have used the $3+1$ relations
$u^\phi=\Gamma (v^\phi +\omega/\alpha)$, $u^t=\Gamma/\alpha$, and
we have defined $\Omega:=u^\phi/u^t=d\phi/dt$, the angular velocity
of the fluid seen by an observer at rest at infinity (a function
of $r$ and $\theta$ for a differentially rotating NS).
The equilibrium condition we are looking for can be directly
derived from the $3+1$ conservative form of the GRMHD equations,
in which the only non vanishing contribution is due to the two poloidal 
components of the momentum conservation in Eq.~(\ref{eq:dudt})
\begin{multline}
\gamma^{-1/2} \partial_j [\gamma^{1/2}\alpha (p +\textstyle{\frac{1}{2}} B^2)] -
\alpha (p +\textstyle{\frac{1}{2}} B^2)\textstyle{\frac{1}{2}}\gamma^{ii}\partial_j\gamma_{ii} \\
- \alpha (\rho h \Gamma^2 v^2 - B^2) \textstyle{\frac{1}{2}} R^{-2}\partial_j R^2 
+ \rho h \Gamma^2 Rv \,\partial_j\omega  \\
+ (\rho h \Gamma^2 - p + \textstyle{\frac{1}{2}}B^2)\partial_j\alpha=0, 
\end{multline}
where $j$ is either $r$ or $\theta$ and we remind that the electric field
vanishes. Recalling that 
$\textstyle{\frac{1}{2}}\gamma^{ii}\partial_j\gamma_{ii}=\gamma^{-1/2} \partial_j \gamma^{1/2}$
and differentiating the definition of $v$, after a few algebraic steps the
following condition can be found:
\be
\frac{\partial_j p}{\rho h} + \partial_j \ln\alpha - \partial_j \ln\Gamma
+\frac{\Gamma^2 v^2\partial_j\Omega}{\Omega-\omega} + 
\frac{\partial_j (\alpha^2R^2B^2)}{2\alpha^2R^2 \,\rho h} =0.
\ee
Integrability of this equation demands the following conditions:
\begin{itemize}
\item A barotropic EoS, $p = p(\rho h)$. 
The simplest choice and the most common assumption is a polytropic law
 \be
 p = K \rho^{1+1/n} \Rightarrow h = 1 + (n+1) K \rho^{1/n},
\ee
where $n$ is the polytropic index (the corresponding
adiabatic index is $\gamma = 1+1/n$).
\item The quantity $F:=u_\phi u^t \equiv \Gamma^2v^2/(\Omega-\omega)$,
related to the specific angular momentum $\ell:=-u_\phi/u_t$, is a function
of the angular velocity $\Omega$ alone. A commonly adopted differential rotation 
law \citep[e.g.][]{stergioulas03} is
\be
F(\Omega)=A^2(\Omega_c-\Omega)=
\frac{R^2(\Omega-\omega)}{\alpha^2-R^2(\Omega-\omega)^2},
\label{eq:Omega}
\ee
where $\Omega_c$ is the central angular velocity and $A$ is a measure of the 
differential rotation rate. For uniform rotators $\Omega\equiv\Omega_c$ (and $A \to \infty$),
and this contribution can be excluded from the Bernoulli integral.
\item A sort of magnetic barotropic law, where $\alpha R B$ is a function of
$\alpha^2R^2 \rho h$. The simplest choice is a magnetic polytropic law 
\be
\alpha R B= K_m(\alpha^2 R^2 \rho h)^m,
\label{eq:B}
\ee
where $m$ is the magnetic polytropic index, with $m \ge 1$ \citep{kiuchi08a}.
\end{itemize}
Using the above prescriptions we can easily derive the final
GRMHD Bernoulli integral
\be
\ln\frac{h}{h_c} + \ln\frac{\alpha}{\alpha_c} -\ln\Gamma 
-\frac{A^2}{2}(\Omega_c-\Omega)^2 + 
\frac{m K_m^2}{2m-1}(\alpha^2R^2 \rho h)^{2m-1} \! = \! 0,
\label{eq:bern}
\ee
where again with the subscript $_c$ we indicate values at the center.

The above derivation is exact for quasi-isotropic coordinates and
obviously applies also to the CFC sub-case. In the non-magnetized
case, it has been demonstrated that, even for rapid rotators close to 
the mass shedding limit (see Sect.~\ref{sect:rns}), solutions obtained
with the RNS numerical code \citep{stergioulas95} show very
little deviations from a CFC metric, so in principle one would
expect the CFC limit to provide a reasonably good approximation of the
correct solution for compact relativistic stars. Given the above mentioned
computational advantages in the solution of the XCFC system of
equations with respect to the original set of coupled PDE of fully
constrained schemes, our choice has been to build a novel
numerical code (written in \verb|Fortran90|), which we name XNS 
and that can be freely downloaded at 
\\ \\
{\small \verb|http://sites.google.com/site/niccolobucciantini/xns|}
\\ \\
This takes advantage of the same XCFC metric 
solvers developed for the X-ECHO scheme described in the
present section.
A comparison between equilibria found with XNS and RNS
is presented in Sect.~\ref{sect:rns}, together with a discussion of
the results.

XNS employs a self-consistent method in the search for the
axisymmetric equilibrium solutions of relativistic compact stars, 
in the presence of differential rotation and a purely
toroidal magnetic field, thus metric terms and fluid-like
quantities are derived at the same time. Given the values of
the six free parameters $K, n, A, \Omega_c, K_m, m$, plus a guess
for the central density $\rho_c$, the following steps are taken:
\begin{itemize}
\item A starting guess for the CFC metric terms $(\alpha,\psi,\omega)$
is provided from the previous step, with $R$ given by Eq.~(\ref{eq:R}). 
The first time, the spherically
symmetric Tolman-Oppenheimer-Volkoff (TOV) solution in isotropic
coordinates \citep{tolman34}, for the metric of a non-rotating and 
non-magnetized  NS with a central density value $\rho_c$, is computed 
through a shooting method for ODEs.
\item On top of these metric terms and for each grid point, $\Omega$ is
derived by inverting Eq.~(\ref{eq:Omega}), then $v$ (and $\Gamma$)
can be determined from Eq.~(\ref{eq:v}). Finally, $h_c$ is a known
function of $\rho_c$ and the Bernoulli integral in Eq.~(\ref{eq:bern}) is
solved via a Newton method to find the local values of $h$ and $\rho$,
allowing us to determine also the magnetic field strength from Eq.~(\ref{eq:B}).
\item Combining the updated fluid quantities with the old metric, the new
conserved quantities $\Ebar, \Sbar_j$ are derived.
\item The set of XCFC equations is solved for this set of conserved
variables, and a new metric is computed. Here only the azimuthal
components for the vector Poisson equations are treated, in order
to find $W^\phi$ and then $\beta^\phi$. We remind here that
the conservative to primitive variables inversion must be enforced
between the solutions to the two scalar Poisson-like equations, namely
after the new value of $\psi$ is found and before that of $\alpha$.
\end{itemize}
These steps are repeated until convergence to a desired tolerance
is achieved. However, a word of caution is in order here. Since XNS is based
on the XCFC metric solver, that works on the conservative variables
densitized with $\psi^6$, convergence is actually enforced on the
central value of the quantity $\hat{D}:=\psi^6D=\psi^6\,\rho_c$.
Therefore, given that  the final conformal factor $\psi$ 
for the self-consistent 2D equilibrium may be quite different from
that derived from the radial TOV solution at the first step, the
final value of $\rho$ at the center is expected to differ from 
the parameter $\rho_c$ in the Bernoulli integral.
If one wants to find an equilibrium converging exactly to a 
central density $\rho_c$, an additional overall iterative loop is needed.

Before concluding the section, some remarks are in order.
One might question the choice of a purely toroidal field versus a
more realistic configuration including a poloidal component. However,
equilibrium with poloidal fields is only possible for uniform
rotators, with magnetic field fully confined within the star. It is
well known \citep{goldreich69} that any poloidal magnetic
field extending outside a rotating NS will lead eventually to a
spin-down of the same, even for a dipole aligned with the rotation
axis, unless the star is surrounded by a true vacuum.
A small charge density of order of $10^{20}\mathrm{cm}^{-3}$ pairs
is enough to break this assumption, even in the case of millisecond rotators 
with $B\sim 10^{17}$~G, and the timescale to replenish an evacuated
magnetosphere is of order of the rotation period. 
For such strong magnetic fields, affecting the overall equilibrium of the NS
and providing a non negligible contribution to the global stress-energy tensor
in the Einstein equations (lower fields have little dynamical effects 
and can be easily treated as perturbations) the problem  becomes
particularly severe because the typical spin-down time for millisecond 
rotators can be as short as $50$~ms. 
Weaker magnetic field can lead to much longer spin-down times,
and  those configurations might be considered {\it quasi}-equilibrium
cases, at least for the time of a typical numerical run (a few
thousands $M$). However they are of little interest, as stated above.

One might also question if purely toroidal configurations are stable,
and, if not, what is the growth rate of instabilities.
A recent study of the stability properties of neutron star with
strong toroidal field has been presented by \citet{kiuchi08b}. 
Their results show that non-rotating neutron stars with a magnetic
polytropic law with $m \ge 2$ are always unstable, and that 
(uniformly) rotating systems are stable only if their kinetic
energy exceed by at least a factor 5 the magnetic energy. 
However, even if their conclusions about stability are supported by 
numerical simulations, it must be pointed out that their criterion is derived 
analytically only in the Newtonian regime, where magnetic field and 
spacetime metric are not coupled. Moreover, their full GR 
results do not consider intermediate cases with $1<m<2$,
thus further investigation is certainly needed.


\section{Numerical results}
\label{sect:tests}

We present here a set of numerical tests of the X-ECHO scheme. Standard
HD/MHD tests in a static background metric (Cowling approximation) have
been already presented elsewhere both for a flat metric 
\citep{delzanna02,delzanna03} and for a given stationary curved spacetime 
\citep{delzanna07}. As discussed in the introduction, the performances 
of the ECHO code in astrophysical scenarios involving a variety of different
conditions like strong shocks, relativistic outflows, strong gravity, and highly 
magnetized systems, have also been already assessed in numerous papers.
For these reasons, the results presented here focus mostly on evaluating 
the quality of our novel metric solver, and the performances of the HD/MHD 
ECHO algorithms when coupled with a dynamical spacetime.
Moreover, the original ECHO scheme was designed for high-order
accuracy, whereas the metric solver implemented in X-ECHO is formally only
second order in space. Given that the performances of high-order
reconstruction techniques has already been tested \citep{delzanna07}, 
for simplicity, we have decided to limit our set-up to second order accuracy, 
both in space and time, which is always a good compromise between
efficiency, accuracy, and robustness.

As discussed in the introduction, only in recent years stable numerical schemes 
for GRMHD in dynamical spacetimes have started to appear. This, together
with the intrinsic degrees of freedom in the choice of gauge and
coordinate systems, typical of GR, have resulted in a lack of well defined, 
agreed upon, standard numerical tests (in the spirit of what shock-tube problems 
are in flat spacetime or accretion/outflows solutions are for a stationary
curved metric). While in 1D a few problems have emerged as standard
benchmarks, this cannot be said about multidimensional cases yet, also
because of a lack of analytical solutions for fully multidimensional
dynamical problems. Some of the tests that we have selected here,
have been already published in the literature, using different numerical schemes, 
both fully constrained \citep{dimmelmeier02, stergioulas04,dimmelmeier06, 
cordero-carrion09}  and hyperbolic \citep{font02, bernuzzi10}, so that we can validate 
our code against existing results. Some other have been done in the perturbative
regime and compared with the results of linear theory, and when possible 
also with previous numerical simulations. However, in an attempt to
evaluate the performances under different conditions, we also present
some novel cases.

Unless otherwise stated, in all our numerical tests we will use a
Courant number of 0.4, an ideal gas EoS with $\gamma = 2$ 
(corresponding to a polytropic index $n=1$ for the initial data in XNS), 
and we will solve the full GRMHD system, including the equation for the 
total energy density $E$, often neglected in isentropic tests for a given
polytropic law.
For the approximate Riemann solver, we use here for the first time the HLLC
solver by \citep{mignone06}, never applied before to GRMHD studies in
curved, evolving spacetimes, to our knowledge. This choice is imposed
by the sharp transition between the rotating NS and the external atmosphere,
since the two-wave solver HLL is found to be too diffusive on the contact
discontinuity.

Given that the astrophysical problems we are mostly interested in, and
toward which the X-ECHO scheme has been developed, involve the ability of
simultaneously handle the high density NS and any low density
outflow/atmosphere/magnetosphere that might surround it, in almost
all of our tests we have included in the domain an extended
atmosphere, which is left free to evolve and respond to the evolution
of the NS. We are aware that in the literature a common practice is to
reset floor values outside the NS, but we believe that this procedure
may in principle lead to violations of conservation properties of the scheme 
at the NS surface. For the same reasons, as stated above, simulations have been 
performed using a ideal gas EoS, which allows us to handle systems where matter in
different thermal conditions (a cold NS versus a hot atmosphere) is present. 

Reconstruction at cell boundaries is achieved for simplicity through
a \emph{monotonized-central} (MC) algorithm, though the other choices
described in the appendix of \citep{delzanna07} are also possible.
The XCFC metric solver is invoked every 10 steps of the HD/MHD 
Runge-Kutta evolution scheme. In 2D runs we use 50 
Gaussian quadrature points and 20 spherical harmonics. 
With these settings, we have managed to make comparable the 
computational times taken by the metric solver, usually slower, and by a single 
Runge-Kutta cycle of the fluid solver. Thus, solving XCFC at every fluid
step will only double the overall times, but no significant improvement
has been noticed in the results.
Finally, grid spacing will be constant both along the radial direction $r$
and the polar angle $\theta$, so the number of points is enough to
specify the grid in each direction.


\subsection{Stability of a TOV stable radial solution}
\label{sect:1dtov}


\begin{figure}[t]
\resizebox{\hsize}{!}{\includegraphics{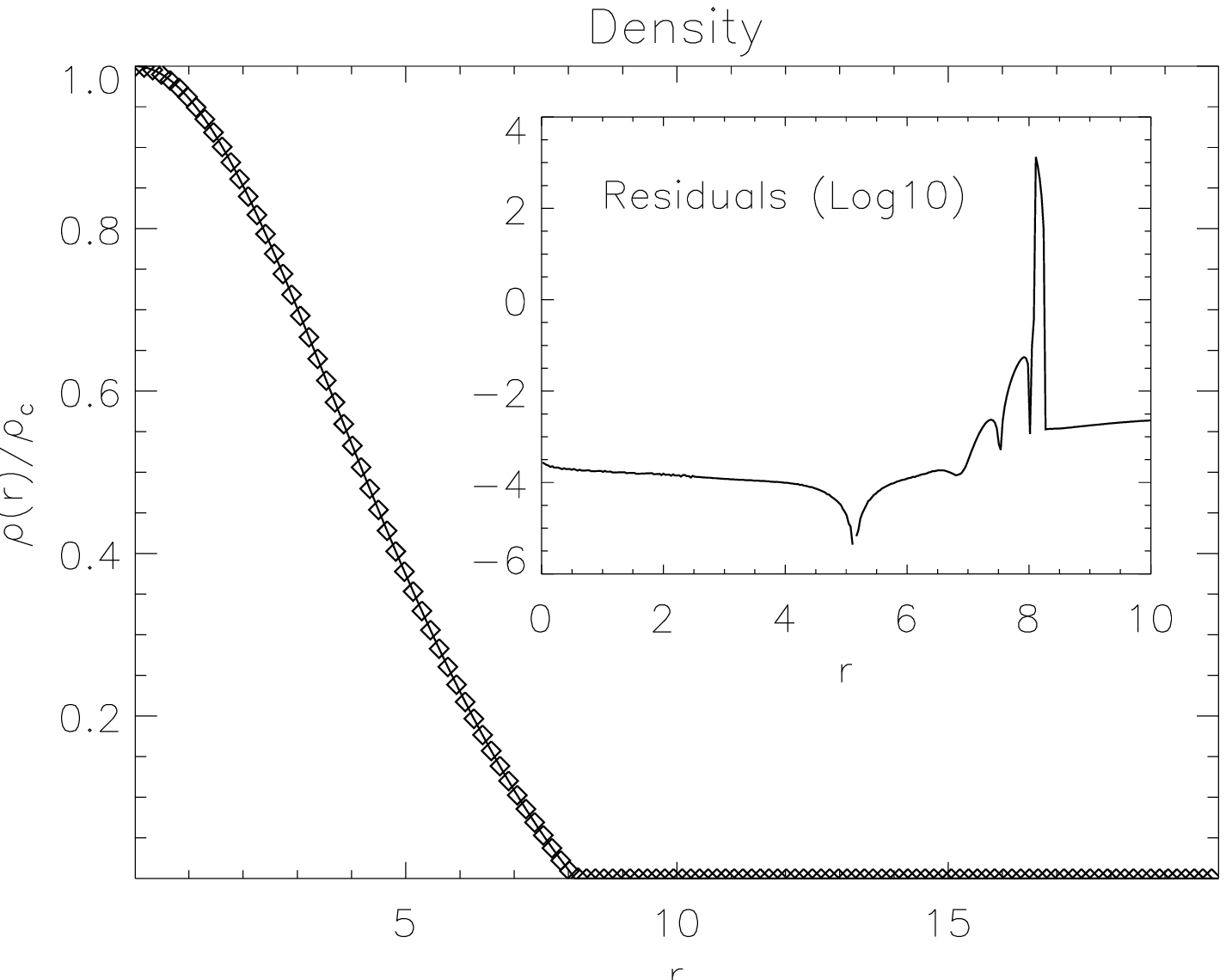}}\\
\resizebox{\hsize}{!}{\includegraphics{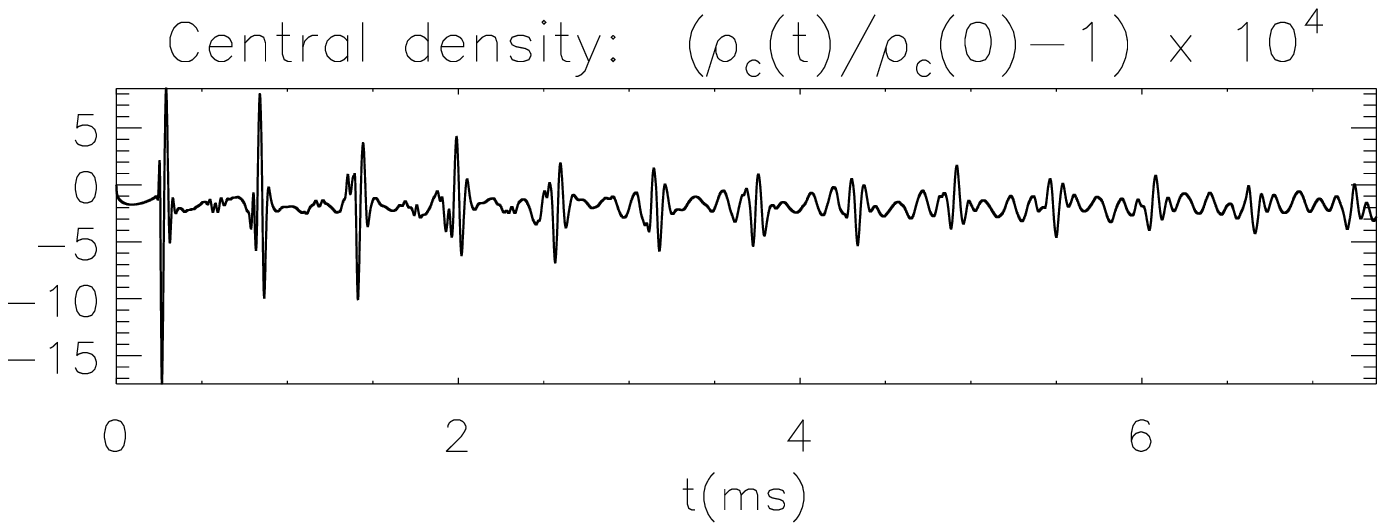}} \\
\resizebox{\hsize}{!}{\includegraphics{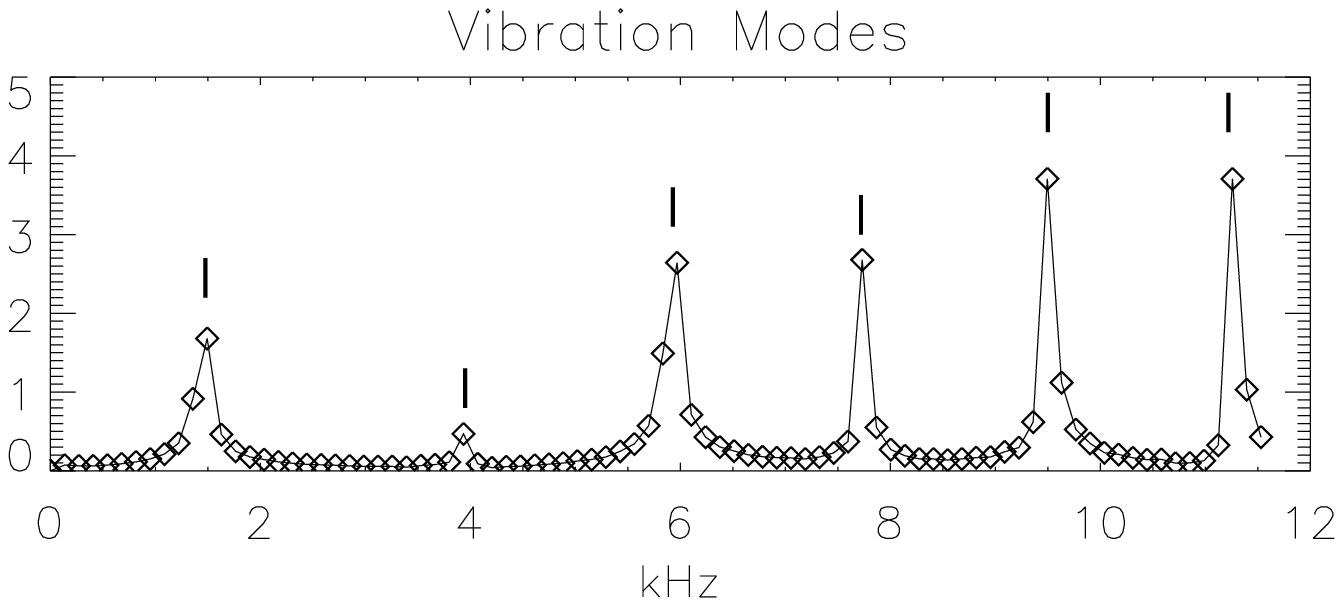}}
\caption{
  Evolution of a stable TOV solution in spherical symmetry and
  isotropic coordinates. The upper panel shows a
  comparison between density in the initial solution (solid line) and the result
  at $t_\mathrm{max}=1500$ (diamonds). For clarity the result a $t_\mathrm{max}$ 
  is shown every 5 points. The insert shows the residuals. The spike at $r\approx 8$ 
  is due to diffusive relaxation at the NS boundary. The middle panel shows the relative 
  variations in time of the central density. The lower panel shows the
  Fourier transform of the the central density. Solid line  and
  diamonds indicate the power of the Fourier series in arbitrary units. The vertical
  markers indicate the frequency of known eigenmodes. The frequency
  resolution of our time series is $\sim 150$~Hz. 
  }
\label{fig:tov1d}
\end{figure} 


Our first test consists in the evolution of a stable 1D radial NS configuration. 
We adopt, as initial condition, a solution of the TOV equations in isotropic 
coordinates, corresponding to a polytropic gas with $K=100, n=1$, and
central density  $\rho_c = 1.280 \times 10^{-3}$, a model also known as A0 (AU0) 
or B0 (BU0) in the literature \citep{font00, stergioulas04, dimmelmeier06}. 
This corresponds to a NS extending to a radius $r = 8.13$. 
It is possible to show that this star lies on the stable part of the
mass-radius curve, so we expect the code to be able to maintain this
configurations for times longer that their typical sound crossing time
($\sim 0.5$~ms). Outside the star we assume, at the beginning of the run, 
a low density and hot ($\rho \simeq 10^{-7}$, $p/\rho\simeq 0.2$) 
atmosphere in hydrostatic equilibrium ($\alpha h = \mathrm{const}$), 
in pressure balance at the surface of the NS. 
Contrary to previous treatments, where the atmosphere was
reset at every time step to keep it stationary, we leave it free to
evolve (collapse or expand) in response to the NS oscillations. 
Given its low density, the atmosphere has negligible feedback on the star.
The simulation is performed using 625 grid points in the radial domain 
$r=[0,20]$, corresponding to a star resolved over 250 points. 
The evolution is followed for a time $t_\mathrm{max} = 1500$ corresponding 
in physical units to $\simeq 7.5$~ms.   

Fig.~\ref{fig:tov1d} shows a comparison between the initial density
profile and that at $t_\mathrm{max}$, together with a plot of the central density 
$\rho_c$ as a function of time. Relative variations of density in the NS interior
are of order of $10^{-4}$,  with major deviations only at the contact discontinuity of
the NS surface, due to diffusion over the much lower density atmosphere.
This triggers the natural vibration modes of the NS, that are the observed
fluctuations, which are the natural outcome for this physical system.
Notice that the central density, plotted in the insert of Fig.~\ref{fig:tov1d},
shows fluctuations of order of $10^{-3}$ at most, but no sign of any secular trend.
This is due to the large number of points over which the star is resolved. 
The slow damping of the oscillations, from $10^{-3}$ to a few $10^{-4}$,
 is due to the thermal dissipation associated to the use of an 
ideal gas EoS and to the numerical viscosity of the scheme, whereas
it is not present if a polytropic EoS is used (see next
Sect. ~\ref{sect:migration} for a comparison between the two EoS). 
At the beginning of the simulation we observe a relaxation of the  central density 
to a value which is $\sim 2\times 10^{-4}$ lower than the initial condition, 
probably because of discretization errors. However the average value seems 
to remain constant at later times. This shows the ability of the code to maintain 
a stable equilibrium, even for several ($\sim 10$) sound crossing times. 
It also shows that the presence of a dynamical atmosphere causes no
problem for the stability of the TOV solution. On the contrary, the
atmosphere itself seems to be quite stable, as shown by the fact that
its density changes by a factor smaller than $1\%$.

In the bottom panel of Fig.~\ref{fig:tov1d} we plot a Fourier transform of the central 
density in time.  Markers indicate the positions of the known eigenmodes. 
This is a test of the performance of the code in handling a dynamical spacetime, at
least in the linear regime, for small perturbations. The values of the
eigenmodes, the fundamental in particular, are quite different in the Cowling
approximation where the metric is kept fixed in time \citep{font02}. 
It is interesting to note also that no initial perturbation has been introduced,
and that the oscillations of the star are only due to the relaxation of
the initial conditions and possible round off-errors.  
Indeed, the large power present in the higher frequency modes
suggests that the initial excitation is confined to small scales, 
most likely at the surface of the star. The presence of a freely evolving
atmosphere does not seem to affect the frequency of the modes, at
least within the accuracy of our temporal series.  


\subsection{Migration of an unstable TOV radial solution}
\label{sect:migration}

A genuinely non-trivial situation in the fully non-linear regime,
involving large variations of the metric and fluid structure, is the
migration of an unstable 1D TOV solution. Following
\citep{font02, cordero-carrion09, bernuzzi10} we select
a solution of the TOV equations in isotropic coordinates,
corresponding to a polytropic gas with again $K=100, n=1$, and
central density  $\rho_c = 7.993 \times 10^{-3}$. This corresponds to a
star extending to a radius $r = 4.26$, on the unstable part of the
mass-radius curve. The external atmosphere is set as in the
previous test.
Due to truncation errors and the initial relaxation of the
NS/atmosphere transition, this configuration
migrates to the stable branch. This evolution causes large amplitude
pulsations, and the formation of a shock between the outer mantle and
inner core of the star, where part of the kinetic energy is dissipated
into heat. During the evolution the star expands to quite large radii. 
The run is done using 900 grid points in the radial domain $r=[0,30]$.
A small fraction of the stellar mass (smaller than $\sim 10^{-3}$) is lost at 
the outer radius during the first bounce. The evolution is followed for a time
$t_\mathrm{max} = 1500$ corresponding in physical units to $\simeq 7.5$~ms.


\begin{figure}[t]
\resizebox{\hsize}{!}{\includegraphics{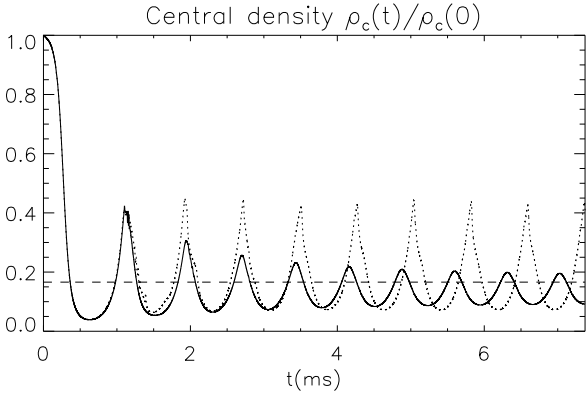}}
\caption{Evolution of the central density for the migrating unstable
  TOV solution in spherical symmetry and isotropic  coordinates: the
  solid line is the evolution for an ideal gas EoS, the dotted line is
  the evolution for a polytropic EoS.
  The horizontal dashed line indicates the density of the stable TOV 
  solution corresponding to the same mass $\rho_c = 1.650 \times 10^{-3}$. }
\label{fig:migration}
\end{figure} 


Fig.~\ref{fig:migration} shows the evolution of the central density in time. 
Results agree with what has been previously presented,
both concerning the amplitude of the fluctuations, the value of the
density at the first minimum and at the first and second maxima,
the frequency of the oscillations,
their non-sinusoidal shape, and the asymptotic value of the central
density with respect to the expected value for a stable configuration with
the same mass. As already noted \citep{font02}, the lower average central 
density at later time (with respect to that of a stable model with the same mass), 
is due to shock heating during the bounce phase, that changes the 
thermal content of the star. We have repeated the same simulation
using a polytropic EoS for the NS (while the ideal gas EoS is still used
for the free atmosphere). Results are shown in Fig.~\ref{fig:migration}
(dotted line). Is is evident that for a polytropic EoS the
oscillations are not damped, and the system appears to converge to the
correct asymptotic value of the central density. Comparison with
previous results \citep{font02,cordero-carrion09} agrees both in the
amplitude of the oscillations and in their phase shift with respect to the
case with an ideal gas EoS.


\subsection{Accuracy of uniformly rotating 2D equilibria}
\label{sect:rns}

Following Sect.~\ref{sect:xns}, we have built a series of 2D equilibrium  
models for rigidly rotating neutron star in the XCFC approximation for the
metric. The computational grid covers the domain $r=[0,20]$ and
$\theta=[0,\pi]$, with 650 zones in radius and 100 zones in angle.
In this section we compare them with equivalent models built
using the publicly available code RNS
\citep{stergioulas95, nozawa98, stergioulas03}, which for us
constitutes a reference benchmark. RNS is an accurate solver
for non-magnetized, uniformly rotating NS configurations
in quasi-isotropic coordinates, for which we recall that
in 2D $R^2\neq \psi^4 r^2 \sin^2\theta$ in Eq.~(\ref{eq:qisotropic}).
This is both a test of the quality of the XCFC approximation, for
genuinely non-spherically symmetric systems, and an evaluation of the
performances of our metric solver. We have selected the BU series of models
from the literature \citep{stergioulas04, dimmelmeier06}, with a fixed central density
$\rho_c=1.280 \times 10^{-3}$, the usual polytropic law with $K=100, n=1$,
but different values of the \emph{uniform} rotation rates. Tab.~\ref{tab:1}
presents the models and compares some global properties derived using
RNS with those derived using our implementation of the XCFC
solver in XNS. Results agree within the accuracy of the models
themselves ($\Delta r =0.03$). The model BU9 represents an equilibrium 
at the mass shedding limit, and it is particularly sensitive to accuracy and
round-off errors. Indeed, when solving the Bernoulli condition
Eq.~(\ref{eq:bern}) to derive the matter distribution, in the case
of model BU9, we had to impose the further condition $\rho=0$ for
$r>11.6$, to avoid unbounded solutions. 


\begin{table}[t]
  \caption{Comparison between RNS and XNS for rigidly rotating
    compact stars. Gravitational masses are in units
    of $M_{\sun}$, $r_e$ and $r_p$ are the equatorial and polar radii.}
  \label{tab:1}
  \centering \begin{tabular}{c c c c c c c c}
 \hline
 Model & $\Omega $ & \multicolumn{2}{c}{$M$} & \multicolumn{2}{c}{$r_e$}
 & \multicolumn{2}{c}{$r_p/r_e$}  \\
 & $(\times 10^{-2})$  & RNS & XNS & RNS & XNS & RNS & XNS \\
\hline
BU0 & 0.000  & 1.400 & 1.400 &   8.13 &  8.13 & 1.00 & 1.00 \\
BU1 & 1.075 & 1.432 & 1.433 &  8.33  &  8.34 & 0.95 & 0.95 \\
BU2 & 1.509 & 1.466 & 1.468 &  8.58  &  8.56 & 0.90 & 0.90 \\
BU3 & 1.829 & 1.503 & 1.505 &  8.82  &  8.85 & 0.85 & 0.85 \\
BU4 & 2.084 & 1.543 & 1.455 &  9.13  &  9.15 & 0.80 & 0.80 \\
BU5 & 2.290 & 1.585 & 1.586 &  9.50  &  9.52 & 0.75 & 0.75 \\
BU6 & 2.452 & 1.627 & 1.629 &  9.95  &  9.98 & 0.70 & 0.70 \\
BU7 & 2.569 & 1.666 & 1.667 & 10.51 & 10.53 & 0.65 & 0.65 \\
BU8 & 2.633 & 1.692 & 1.693 & 11.26 & 11.30 & 0.60 & 0.60 \\
BU9 & 2.642 & 1.695 & 1.698 & 11.63 & 11.60 & 0.58 & 0.58 \\
\hline
\end{tabular}
 \end{table}


\begin{figure}
\resizebox{\hsize}{!}{\includegraphics{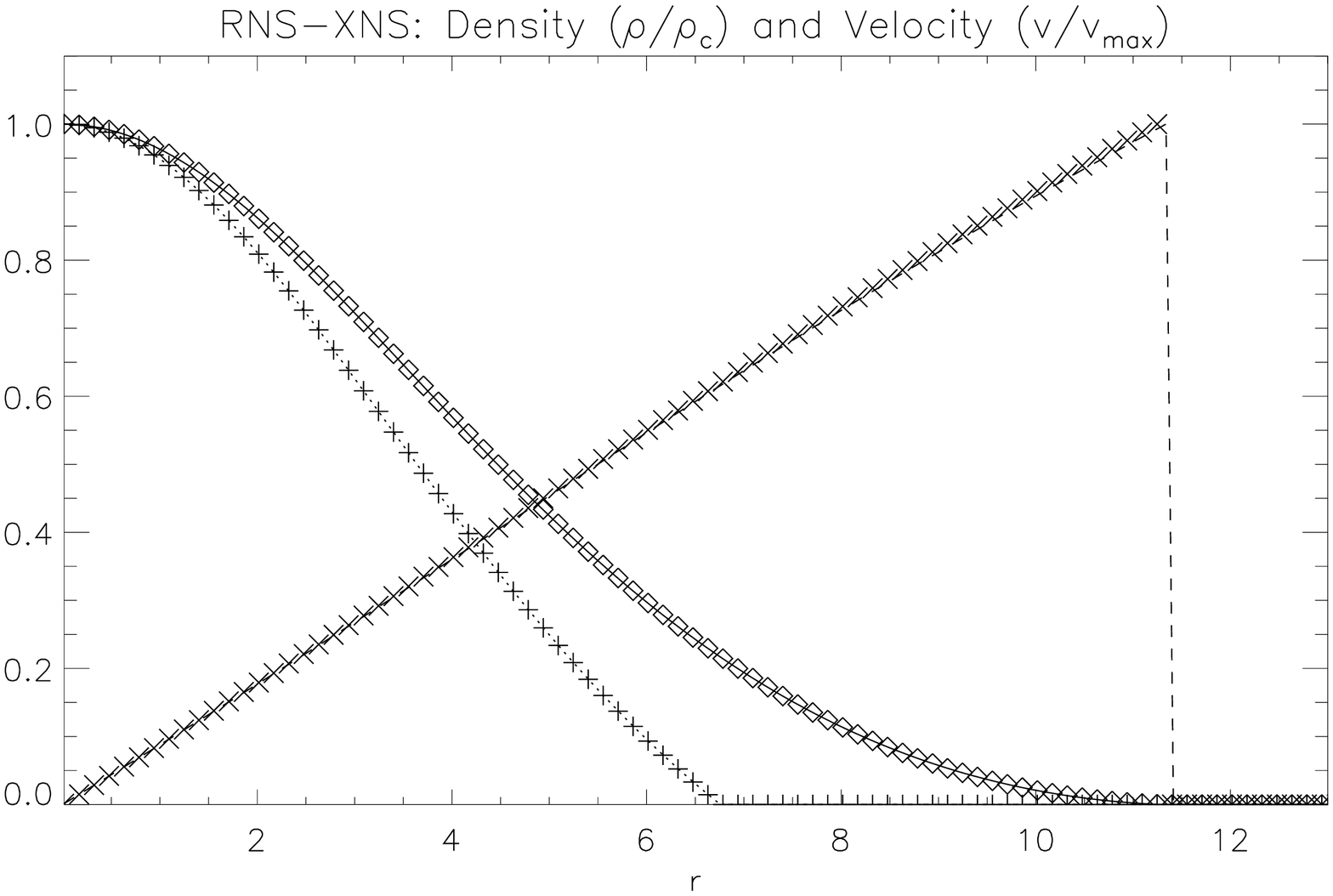}}\\
\resizebox{\hsize}{!}{\includegraphics{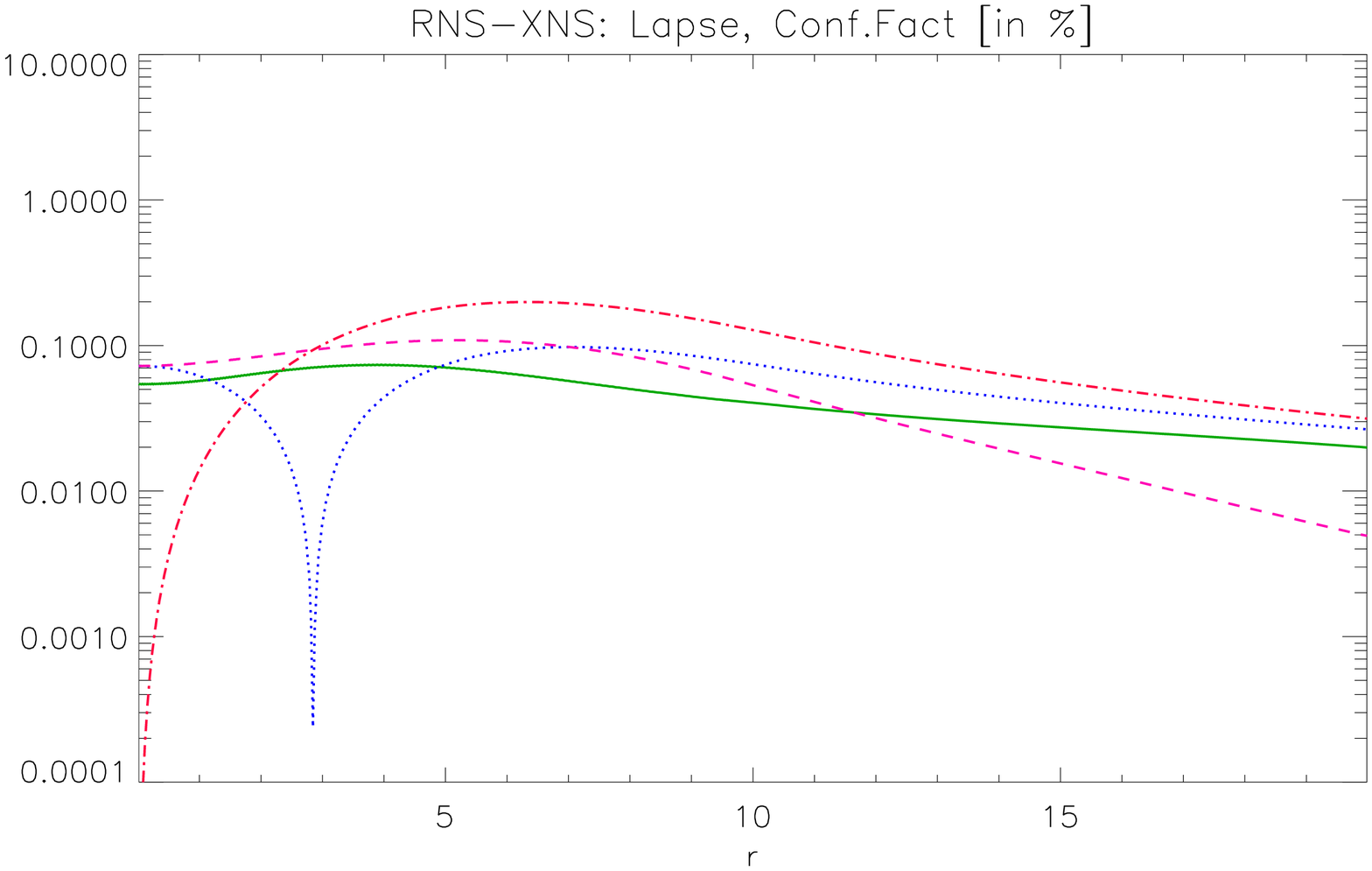}}
\caption{Comparison between RNS and XNS solutions, for model BU8. 
The upper panel shows the fluid quantities. The solid (dotted) line represents
the density at the equator (polar axis) derived using RNS, both normalized
against $\rho_c$. The dashed line is the profile of the rotational velocity 
module $v$, normalized to its maximum ($0.37462$). Diamonds, crosses and pluses
represent values of the same quantities as derived using XNS, where
for clarity we report one symbol every 5 radial points. 
The lower panel shows the residual of various metric terms at the equator. 
The green solid line is the relative error between the lapse $\alpha$ computed
with XNS and RNS. Dashed magenta and dotted blue lines represent the relative error between 
the conformal factor $\psi$ of the CFC metric with respect to that in
quasi-isotropic coordinates and the quantity $[R/(r\sin\theta)]^{1/2}$.
The dot-dashed red line represents the difference between $\psi$ and
$[R/(r\sin\theta)]^{1/2}$ both computed in quasi-isotropic coordinates, 
and can be considered a measure of the non
conformal flatness of the RNS solution.}
\label{fig:rnscomp}
\end{figure} 


In Fig.~\ref{fig:rnscomp} we show the comparison between the model BU8
(we already pointed out the problems for model BU9)
derived using our solver for the CFC metric and the solution of RNS. 
It is clear that the CFC approximation provides a good
description of the matter and fluid properties of the equilibrium
configuration, throughout the entire star. The densities along the polar axis and at
the equator, together with the velocity profile $v:=(v_\phi v^\phi)^{-1/2}$, are
all well reproduced. In terms of the metric coefficient, we see that the discrepancy is of
order $10^{-3}$, and is comparable with the level of non conformally
flatness, defined as the difference between $\psi$ and $[R/(r\sin\theta)]^{1/2}$
in Eq.~(\ref{eq:qisotropic}). Somewhat larger deviations, of order a few 
$10^{-3}$ in the NS interior, are characteristic of the shift $\beta^\phi$,
increasing at larger radii where the value of  the shift approaches zero, 
analogously to what was observed by \cite{dimmelmeier02}. 


\subsection{Stability of uniformly rotating 2D equilibria}
\label{sect:rot2d}

In the same spirit as for the numerical test presented in
Sect.~\ref{sect:1dtov}, we show here the result of a time evolution of
two equilibrium configurations, BU2 and BU8. They correspond to a
mildly rotating star and to a case of rapid rotation, respectively. The initial condition 
are derived according to the recipe given in Sec.~\ref{sect:xns} in conformally
flat metric, and their accuracy has already been discussed in the previous section. 
The domain, $r=[0,16]$ and $\theta=[0,\pi]$,  is spanned by 200 zones 
in the radial direction and 100 zones in angle. 
The star is resolved on average over about 100 radial zones. 
The evolution is followed for a time $t_\mathrm{max}=1500$, 
corresponding in physical units to $\simeq 7.5$~ms.  As in the previous
tests, we initialize, outside the star, a hot and low density atmosphere, 
initially in equilibrium and then allowed to freely evolve in time.
As was already noted in the previous 1D cases, even in 2D the presence of this
freely evolving atmosphere has negligible dynamical effects on the star,
due to its low density, and there is no need of enforcing a reset to the
initial values. 


\begin{figure}[t]
\resizebox{\hsize}{!}{\includegraphics{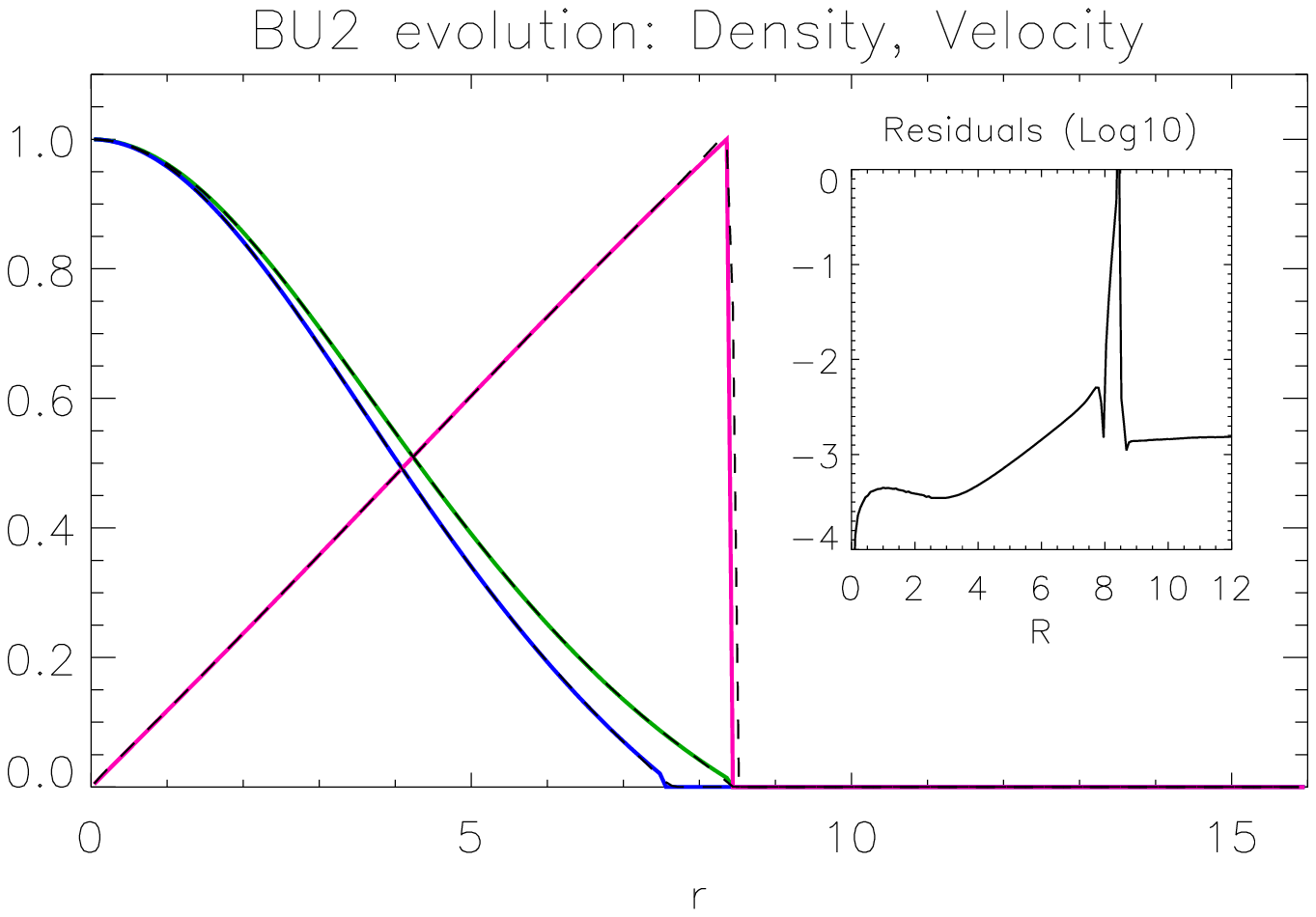}}\\
\resizebox{\hsize}{!}{\includegraphics{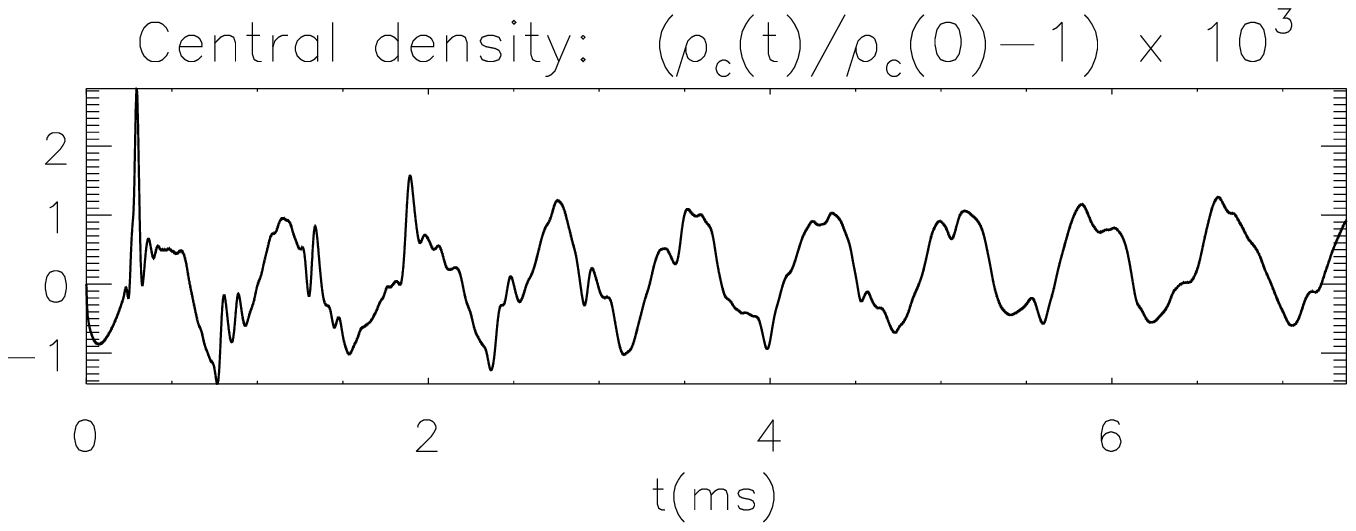}} \\
\resizebox{\hsize}{!}{\includegraphics{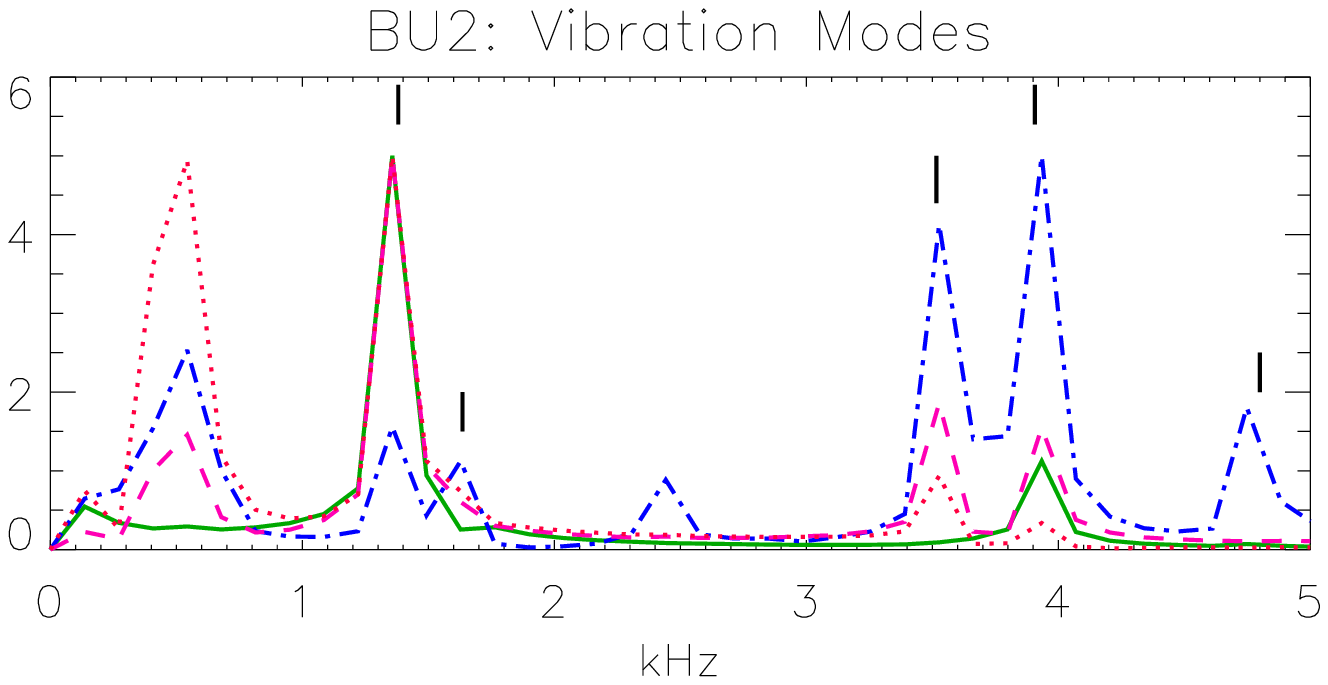}}
\caption{Evolution of a stable BU2 solution. The upper panel shows a
  comparison between the initial values  (solid lines) of equatorial
  (green) and axial (blue) densities, 
  and equatorial rotational velocity $v$ (magenta), against the value 
  of the same quantities at at $t_\mathrm{max}=1500$ (dashed lines). 
  The densities are normalized to the central initial value $\rho_c$, 
  the velocity to its maximum initial value ($0.1619$). 
  The insert shows the relative difference between the equatorial 
  densities as a function of radius. The spike at $r\approx 8$ is due
  to diffusion at the NS surface, also partly visible in the velocity profile. 
  The middle panel shows the variation in time of the central density. 
  Note that there is a hint of a secular trend of increasing density, 
  with an average increase during the whole run of a few $10^{-4}$.
  The bottom panel shows the Fourier transform of the the density
  (green solid line), 
  $v^r$ (red dotted line), $v^\phi$ (magenta dashed line), and
  $v^\theta$ (blue dot-dashed
  line), at the point  $r=3.0$, $\theta=45^\circ$ of model BU2. 
 The vertical markers indicate the frequency of known eigenmodes. 
 The frequency resolution of our time series is $\sim 150$~Hz.
  }
\label{fig:bu2}
\end{figure} 


\begin{figure}[t]
\resizebox{\hsize}{!}{\includegraphics{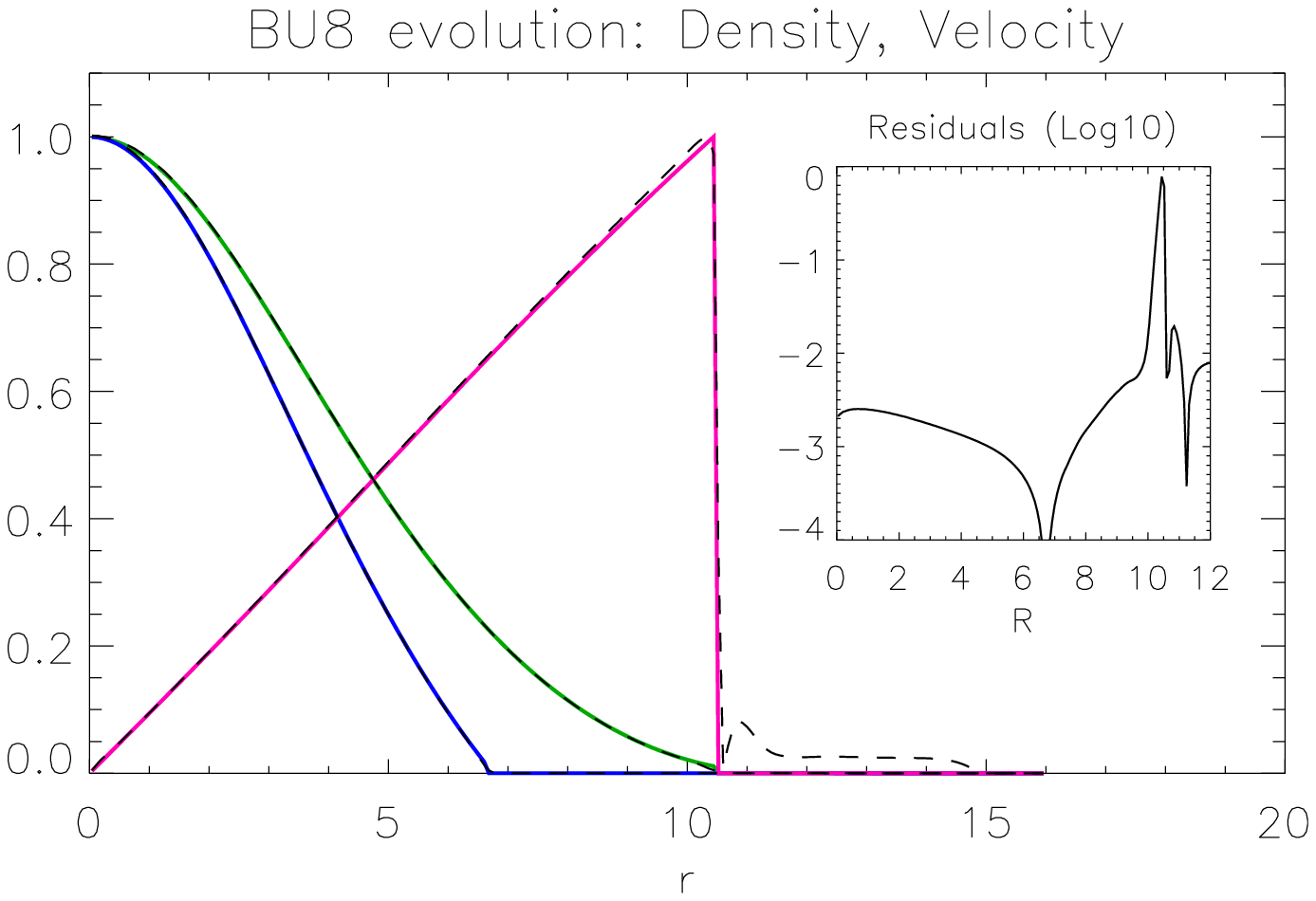}}\\
\resizebox{\hsize}{!}{\includegraphics{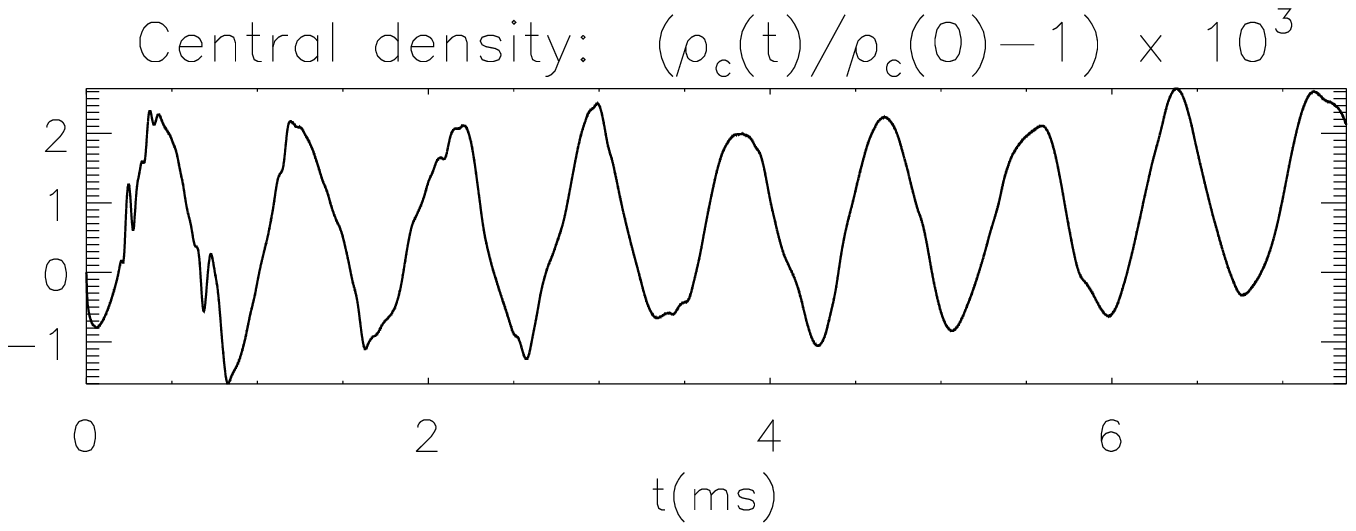}} \\
\resizebox{\hsize}{!}{\includegraphics{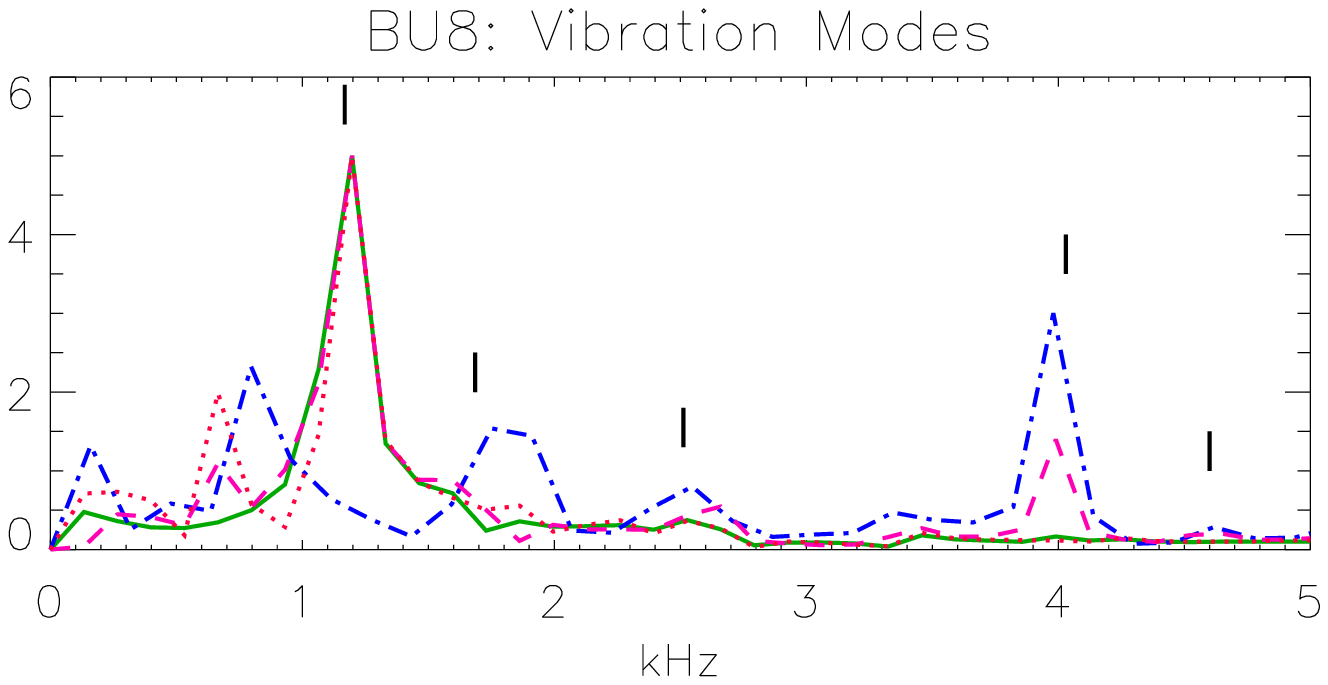}}
\caption{Evolution of a stable BU8 solution. The upper panel shows a
  comparison between the initial values (solid lines)  of equatorial
  (green) and axial (blue) densities, 
  and equatorial rotational velocity $v$ (magenta), against the value 
  of the same quantities at at $t_\mathrm{max}=1500$ (dashed lines). 
  The densities are normalized to the central initial value $\rho_c$, 
  the velocity to its maximum initial value ($0.3499$). 
  The insert shows the relative difference between the equatorial 
  densities as a function of radius. The spike at $r\approx 10$ is due
  to diffusion at the NS surface, also partly visible in the velocity profile. 
  The middle panel shows the variation in time of the central density. 
  Note that there is a hint of a secular trend of increasing density, 
  with an average increase during the whole run of a few $10^{-3}$.
  The bottom panel shows the Fourier transform of the the density
  (green solid line), 
  $v^r$ (red dotted line), $v^\phi$ (magenta dashed line), and
  $v^\theta$ (blue dot-dashed
  line), at the point  $r=3.0$, $\theta=45^\circ$ of model BU8. 
 The vertical markers indicate the frequency of known eigenmodes. 
 The frequency resolution of our time series is $\sim 150$~Hz.
}
\label{fig:bu8}
\end{figure}


For the model BU2, Fig.~\ref{fig:bu2} compares the equatorial and axial
densities together with the module of the rotational velocity $v$, at time
$t=0$ and $t=t_\mathrm{max}$. The evolution of the central density is also
shown. The oscillations of order or $10^{-3}$ are due to the
excitation of the stellar eigenmodes by initial round-off errors and
relaxation at the boundary between the star and the atmosphere. 
No initial perturbation was introduced into the model. Changes in the
value of the equatorial density at the end of the run, with respect to
the initial values, are of the same order of the amplitude of the
oscillations that are excited by the initial relaxation. A small
secular drift of order $\sim 5\times 10^{-4}$ in the average value of
the central density is visible by the end of the run. It is due
to the lower number of points in the radial direction over which the
star is resolved ($\sim 100$), compared to the 1D case ($\sim 250$)
where no drift was observed. In the bottom panel of Fig.~\ref{fig:bu2},
we plot a Fourier transform of various fluid quantities in time. The
quantities are all measured at a selected location inside the star: $r=3.0$ and
$\theta=\pi/4$. Normal mode analysis is beyond the scope of this
paper, so here we just present a simple single-point analysis. Selecting
other points in the star does not change the location of the peaks in
the power series (though it affects their relative amplitude). Given that no
perturbation is introduced in the initial data, modes are differently
excited, depending on the initial relaxation (certain modes can in
principle  even not be excited). Markers
indicate the positions of the known $l=0$, $l=2$ and $l=4$ eigenmodes 
\citep{dimmelmeier06}. This is a test of the performance of the code 
in handling an axisymmetric dynamical spacetime, at
least in the linear regime for small perturbations. The fundamental $F$ and first
overtone $H_1$ of the $l=0$ mode are correctly recovered, together with
the $^2 p_1$ for the dipolar $l=2$ mode, and the $^4 p_1$
for the quadrupolar $l=4$ mode. The $^2 f_1$ dipolar
mode is only visible in the spectrum of $v^\theta$. The
$v^\theta$ spectrum also shows a peak at $2.45$~kHz, which does not
correspond to any of the known low frequency modes. It is possible
that its origin might be due to some form of non-linear coupling, or
perhaps an effect of the free atmosphere. However, as already 
noted in the 1D radial case, the presence of a freely evolving
atmosphere does not seem to affect the frequency of the modes, at
least within the accuracy of our temporal series. 
It is interesting to recall that inertial modes can be also excited
with a continuum spectrum extending in a frequency range
between $0$ and $2\Omega$ \citep{font02}. The peak at $500$~Hz,
corresponds in fact to the rotation frequency of the star.

We have repeated the same analysis for the model BU8, with the same
simulation setup. This represents a rapidly rotating case, close to the
mass shedding limit, and it is a more demanding test than the mildly
rotating case, BU2. Results are shown in
Fig.~\ref{fig:bu8}. Many of the same considerations,
as for the previous BU2 case, still apply and will not be repeated.  
Typical deviations in the value of the density are larger (about a
factor 2) than in the previous case. There is also evidence for a
secular drift (increase) of the central density, whose average value
at the end of the run is a factor $10^{-3}$ higher than at the
beginning. Again this drift seems to be twice what is observed in the
BU2 case, suggesting a drop in accuracy at higher rotation rates.
We also show a Fourier transform of various fluid quantities in time. 
These are all monitored at the same selected location: $r=3.0$ and
$\theta=\pi/4$. It is clear that in this case the first $l=0$ mode $F$
is by far the most strongly excited. However, power in the first $l=0$ overtone
$H_1$, at the quadrupole $l=4$ mode $^4p_1$, and $l=2$ overtone $^2p_1$,
is also present. Little power seems to be present at the $l=2$ $^2f_1$
mode. The time evolution of the $v^\theta$ component is quite noisy:
there is no evidence of power at the $H$ mode frequency,
the peak at $1.7$ kHz might be a poorly resolved $^2f_1$ mode, or a
contaminated inertial mode close to $2\Omega$ frequency. 
The power that is visible around $700-800$~Hz 
is likely due to inertial modes, and, analogously to the BU2
case, its frequency corresponds to the rotation frequency of the star. 


\subsection{Stability of differentially rotating magnetized equilibria}
\label{sect:mrns}

We present here the evolution of an equilibrium configuration, which
is both differentially rotating and contains a strong toroidal magnetic
field. The conditions for such equilibrium and how to build it have been
described in Sec.~\ref{sect:xns}. There are no similar tests presented
in the literature, and, as a consequence, no reference against which to compare 
our results. However, this setup allows us to investigate a strongly
magnetized case, where we expect the magnetic field to have
significant dynamical effects. Tests in the literature have, at most,
focused on the case of weak poloidal fields \citep{cerda-duran08},
for the reasons discussed in Sect.~\ref{sect:xns}. However, we are interested here
is evaluating a case with a magnetic field that is not simply a small
perturbation, but that has enough energy to modify the underlying equilibrium. 
Given that the algorithm described in Sec.~\ref{sect:xns} consider only
toroidal fields, we limit our analysis to such a case and we hope that
the proposed test can become a standard, once our XNS solver
will be made of open use. 

The equilibrium model has the same central density of the non-magnetized 
cases $\rho_c=1.280 \times 10^{-3}$, and the usual polytropic law $K=100, n=1$. 
We adopt a differential rotation profile with
$\Omega_c=0.02575$, $A^2=70$, and a magnetic field characterized
by a magnetic polytropic index $m=1$ and coefficient $K_m=3$,
corresponding to a maximum magnetic field inside the star of $\simeq
5\times 10^{17}$~G, and to a ratio of magnetic energy to total internal 
energy  $\simeq 0.1$ (we name this new model BM). 
With respect to an equilibrium with a very similar rotational structure but no
magnetic field \citep[model B2:][]{stergioulas04, dimmelmeier06}
the star is a factor $\sim 1.5$ larger in equatorial radius.
The domain, $r=[0,16]$ and $\theta=[0,\pi]$, is covered with 200 zones in the radial 
direction and 100 zones in angle. The star is resolved on average over about 120
radial zones. The evolution is followed until a final time $t_\mathrm{max}=1500$ 
corresponding in physical units to $\simeq 7.5$~ms.


\begin{figure}[t]
\resizebox{\hsize}{!}{\includegraphics{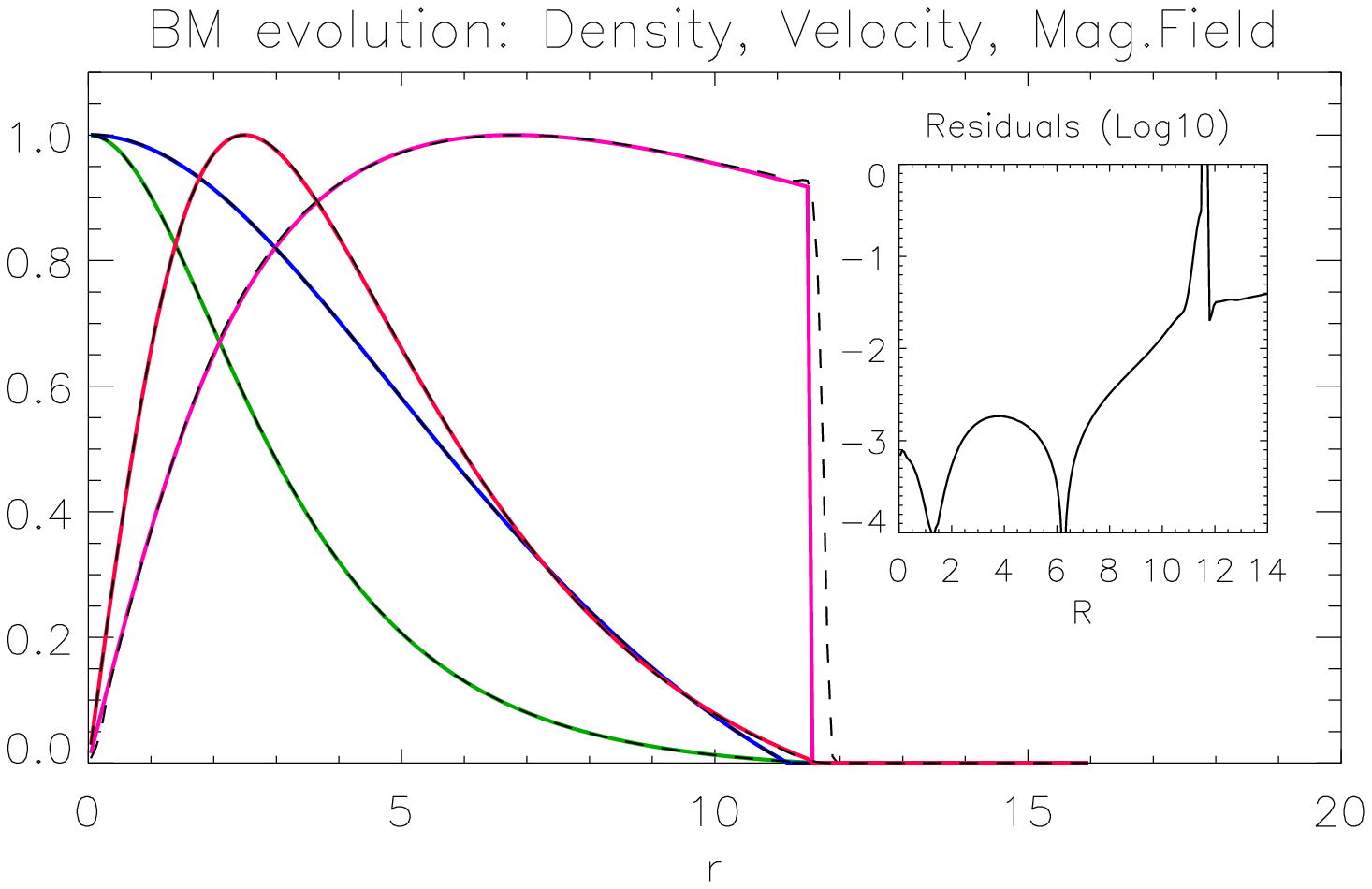}}\\
\resizebox{\hsize}{!}{\includegraphics{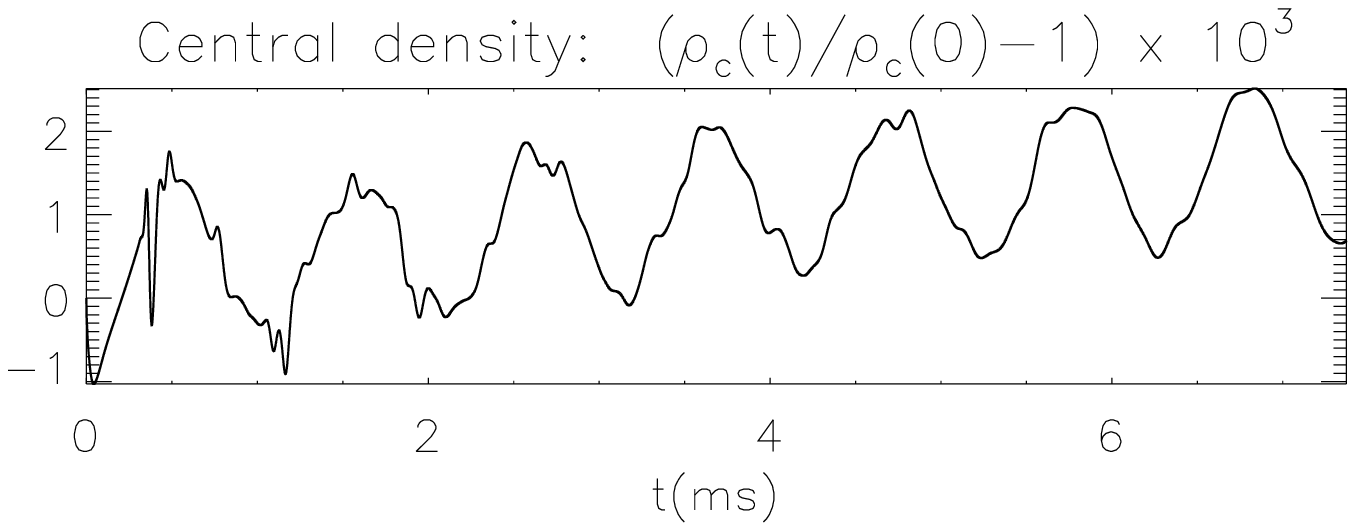}}
\caption{Evolution of a stable magnetized solution. The upper panel shows a
  comparison between the initial profiles (solid lines) of the
  equatorial (green) and axial (blue) densities, 
  the equatorial rotational velocity $v$ (magenta) and toroidal magnetic field $B$ 
  (red), and the value of the same quantities at 
  at t=1500 (dashed lines). Densities are normalized to the initial
  central value, velocity to its maximum ($0.09810$) initial value, and magnetic field 
  to its maximum initial value too. The insert shows the relative difference between
  the equatorial densities as a function of radius. The spike at $r\approx 12$ is due
  to diffusive relaxation at the boundary between the high density star
  and low density atmosphere. The lower panel shown the variation in
  time of the central density, at $\theta=\pi/2$. Note that there is a
 secular trend of increasing density, with an average increase
 during the simulation run of  $\sim 1.5\times 10^{-3}$.}
\label{fig:bm}
\end{figure} 


Fig.~\ref{fig:bm} compares the equatorial and axial profiles of the density,
together with the rotational velocity  $v=(v_\phi v^\phi)^{1/2}$, and the
magnetic field  $B=(B_\phi B^\phi)^{1/2}$, at time
$t=0$ and $t=t_\mathrm{max}$.  There is evidence for a
secular drift (increase) of the central density, whose average value
at the end of the run is a factor $1.5\, 10^{-3}$ higher than at the
beginning, together with typical oscillation of similar
amplitude. Larger deviations of order of $10^{-2}$ characterize the
outer layer of the star for $r>10$, where the ratio of magnetic pressure to
gas pressure is larger. This is also visible in the broader shear
layer that form at the boundary with the atmosphere, compared to non
magnetized cases (Figs.~\ref{fig:bu2}-\ref{fig:bu8}). There is
also some evidence for a drop in the rotation rate of the core. The
reason is not clear, it could either be due to some angular momentum
redistribution from the core to the outer layer of the star, or to
some convection taking place in the core, possibly excited by initial
relaxation. It is likely that, due to round-off errors and to the initial
relaxation, some free energy might be injected into the star to power 
convective motions in regions of marginal convective stability as the core.
We have however verified that the typical poloidal velocities in the core
are $\sim 10^{-4}$, which implies that it will be necessary to follow the
system for a much longer time in order to see whether this convective
motions are stabilized. This also agrees with the results presented by
\citet{kiuchi08b}, which suggest, even for the case of strong
instabilities, typical convective timescales of a few thousands of $M$.

A detailed analysis of the stability of
strongly magnetized configurations is beyond the scope of this paper. 
This test however demonstrates that stable configurations are
preserved for many sound-crossing times, in cases involving strong
magnetic fields too. We have not carried out a full mode investigation, as in
the previous cases, because there are no known values to compare our
results to. There is also no reference solution, as in the case of
RNS for non magnetized rigid rotators. 
However, already from the plot of the central density, it
is possible to see an oscillatory behavior, with a typical frequency
$900$~Hz. The amplitude of the oscillations, $10^{-3}$, is about one
order of magnitude smaller that the result of \citet{kiuchi08b}.
This frequency can be compared with the fundamental mode of a non
magnetized configuration, with a similar velocity profile
($\Omega_c=0.02575$, and $A^2=70$), which is located at $1.37$ kHz
\citep{dimmelmeier06}. There are also some hints that the first $l=0$ overtone 
might have a similarly smaller frequency, even if there is little power in it 
and its identification in the spectrum is not certain. A lower frequency
for $l=0$ modes can easily be understood, if one recall that it is well
known \citep[e.g.][]{bucciantini03} that toroidal magnetic fields behave as gas with
adiabatic index $4/3$, for $l=0$ perturbations. Compared with the gas,
which has adiabatic index $2$, the presence of strong toroidal magnetic field leads to
a softening of the generalized EoS of the perturbations, corresponding to a lower sound
speed, and longer vibrational periods. The larger radial extent of the
star also contributes, given that the period or radial compressive
modes scales as the sound crossing time. 


\subsection{Radial and axisymmetric collapse to a black hole}
\label{sect:bhcollapse}

The migration test performed in Sect.~\ref{sect:migration} already
shows the ability of the XCFC algorithm to handle the dynamical evolution of
very compact configurations. Although that test is failed by the
original CFC \citep{dimmelmeier02}, \citet{marek06} have shown that it can still be
succesfully simulated by using a modified version. The superiority of
XCFC  formulation fully manifests itself in cases where the evolution leads to the
formation of black holes and appparent horizons (AHs)
\citep{york89,baumgarte03,thornburg07}. To properly
evaluate the strength of our XCFC solver, we present in this section
the results of two simulations of the collapse of unstable neutron
stars to black holes: a 1D collapse of a spherically symmetric NS, and
a 2D collapse of a rapidly rotating one
\citep{bernuzzi10,cordero-carrion09,baiotti05}.  In the 1D case the AH
is found using a zero-finding algorithm, while for the 2D collapse we
use a simple minimization algorithm, with the AH parametrized as a surface in terms
of Spherical Harmonics [for details on the algorithms see Sect 8 of
\citet{thornburg07}, Sect 6.7 of \citet{alcubierre08}, and \citet{shibata97}].

For the 1D case we consider the same unstable configuration of the
migration test, Sect.~\ref{sect:migration}, and following
\citet{bernuzzi10} we add a perturbation of the form
\be
\delta v_r = \left\{ 
\begin{array}{l l}
  -0.005 \sin{(r/r_{NS})} & \quad \mbox{if $r \le r_\mathrm{NS}$}\\
   0 & \quad \mbox{if $r > r_\mathrm{NS}$}\end{array} \right. 
\ee
where $r_\mathrm{NS}$ is the initial radius of the neutron star. The
computational domain $r=[0,10]$ is covered by 300
equally spaced radial zones. A hot low density atmosphere is set
outside the NS, and let evolve freely. An ideal gas EoS is used, but
given that the evolution of the collapse does not lead to shocks and
dissipative heating, results are equivalent to the case of a polytropic
EoS. Despite our resolution being 300 times worse than the central
resolution used by \citet{cordero-carrion09}, we are able to follow the
evolution of the system past the formation of an apparent horizon. 


\begin{figure}[t]
\resizebox{\hsize}{!}{\includegraphics{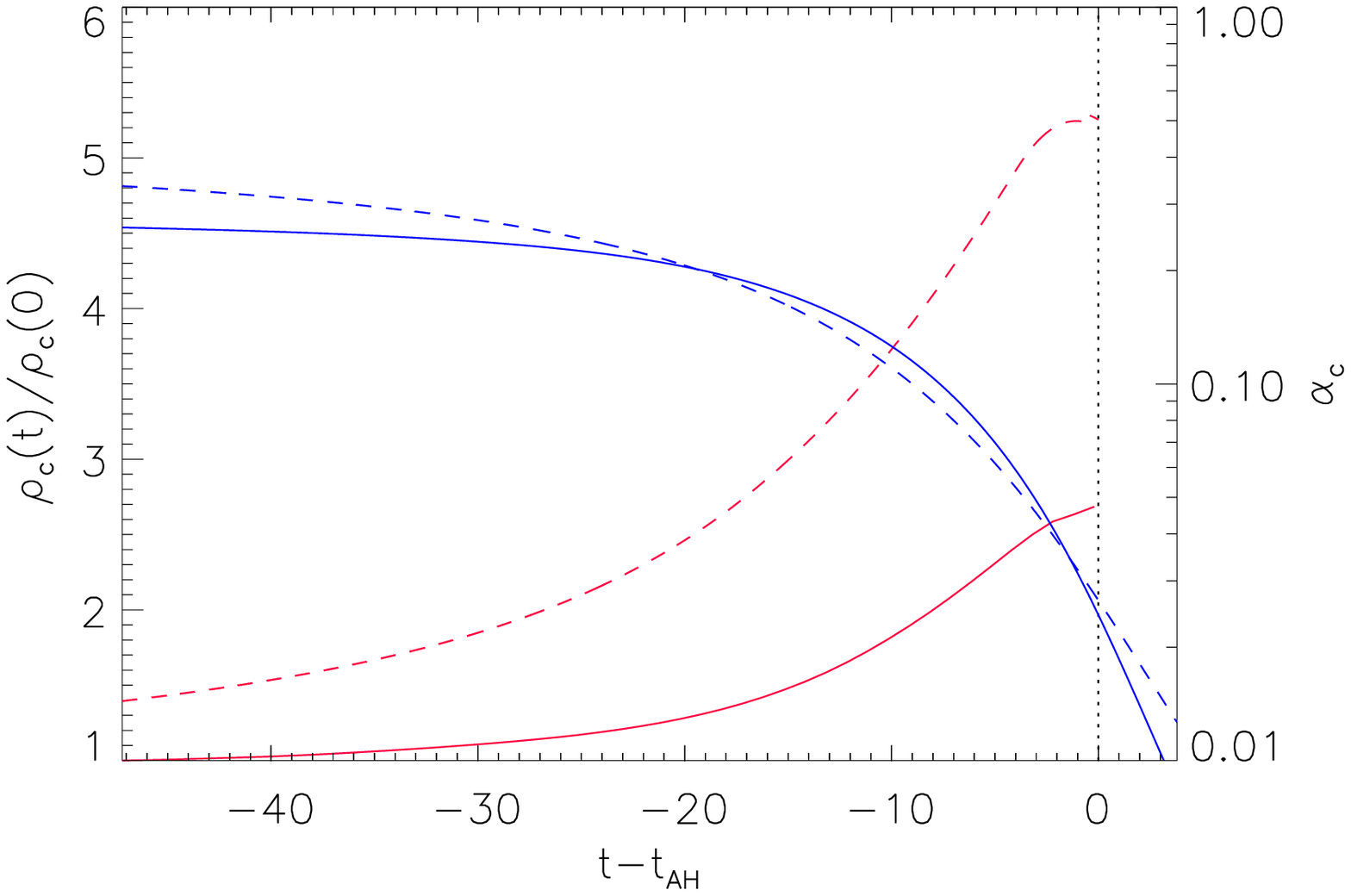}}
\resizebox{\hsize}{!}{\includegraphics{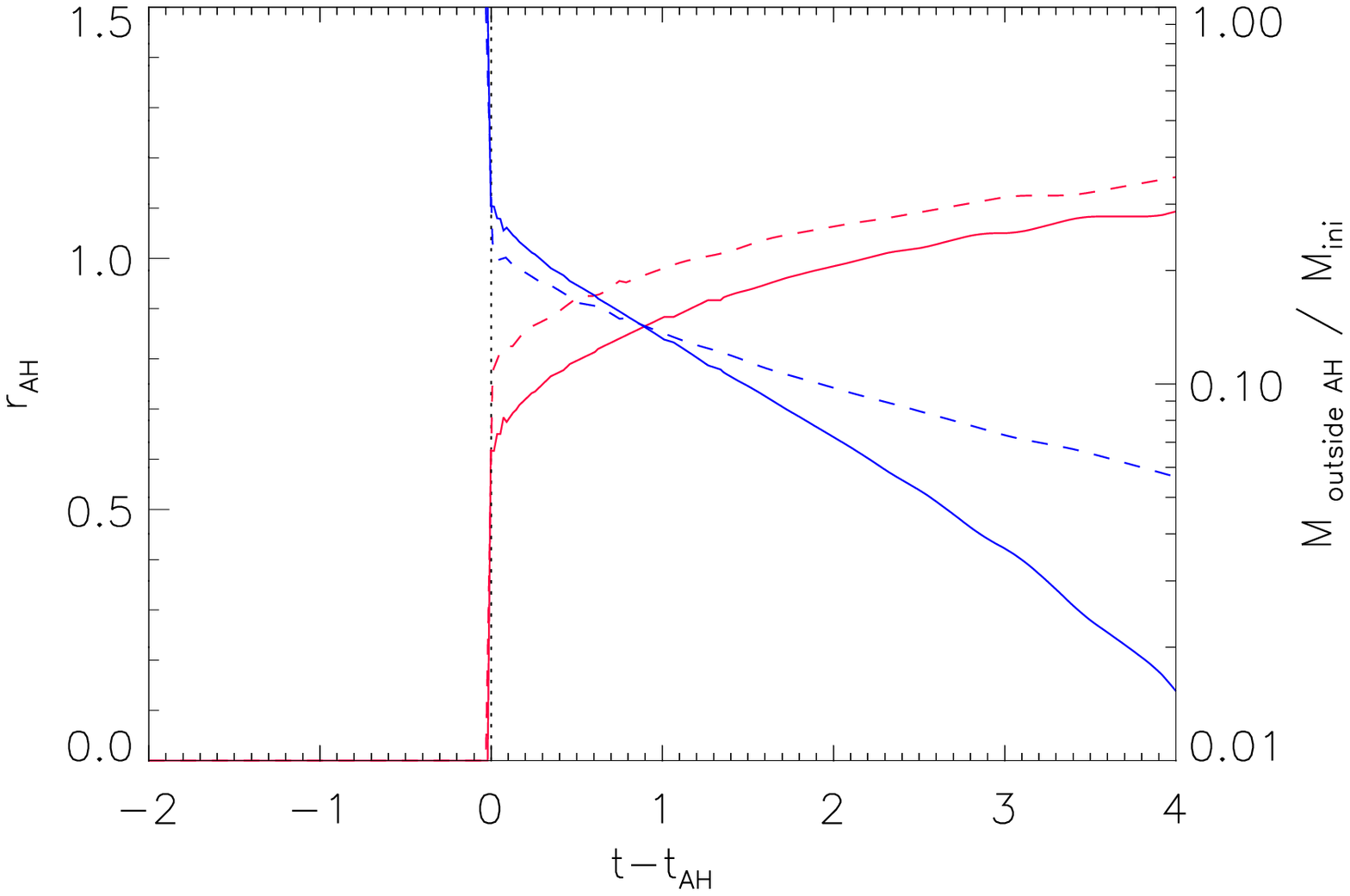}}
\caption{Upper panel: the monotonically rising red curves represent
  the evolution of the central density
  with respect to the intial value $\rho_c/\rho_c(t=0)$ (as explained
  in the text the curves are truncated at the formation of the apparent
  horizon), while the monotonically decreasing
  blue curves represent the value of the lapse $\alpha$ at the
  origin. Solid lines indicate the
  spherically symmetric collapse and dashed lines the collapse of the
  rotating NS model D4. Bottom panel: the rising red curves represent the
  radius of the apparent horizon, while the decreasing blue curves are the
  ratio of the rest mass outside the apparent horizon with respect to
  the total rest mass. Again, solid lines indicate the
  spherically symmetric collapse and dashed lines the collapse of the
  rotating NS model D4. The vertical dotted line
  indicates the time $t_{AH}$ when the apperent horizon forms.   }
\label{fig:bhcollapse1}
\end{figure} 



\begin{figure}[t]
\resizebox{\hsize}{!}{\includegraphics[scale=1.5,bb=20 2 390 450,clip]{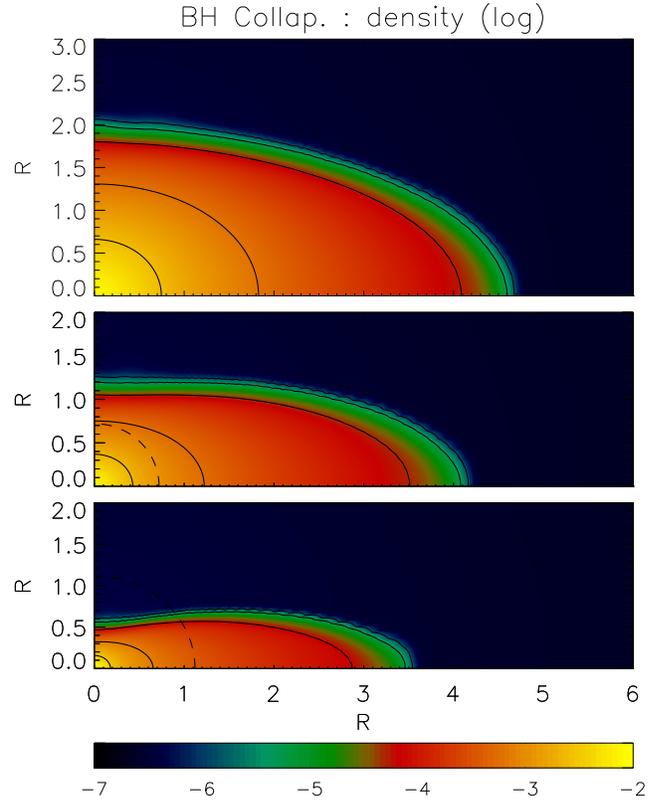}}
\caption{Evolution of the collapse of the rotating unstable
  equilibrium model D4 (density in Log10 units): upper panel, density
  at $t=t_{AH}-6$; middle panel density
  at $t=t_{AH}$; lower panel, density
  at $t=t_{AH}+6$. The dashed countour indicates the position of the
  apparent horizon. At the end of the simulation a left over disk has formed in
  the equatorial region while the neutron star has been completely
  accreted in the polar region.}
\label{fig:bhcollapse2}
\end{figure} 



\begin{figure*}[t]
\resizebox{\hsize}{!}{\includegraphics{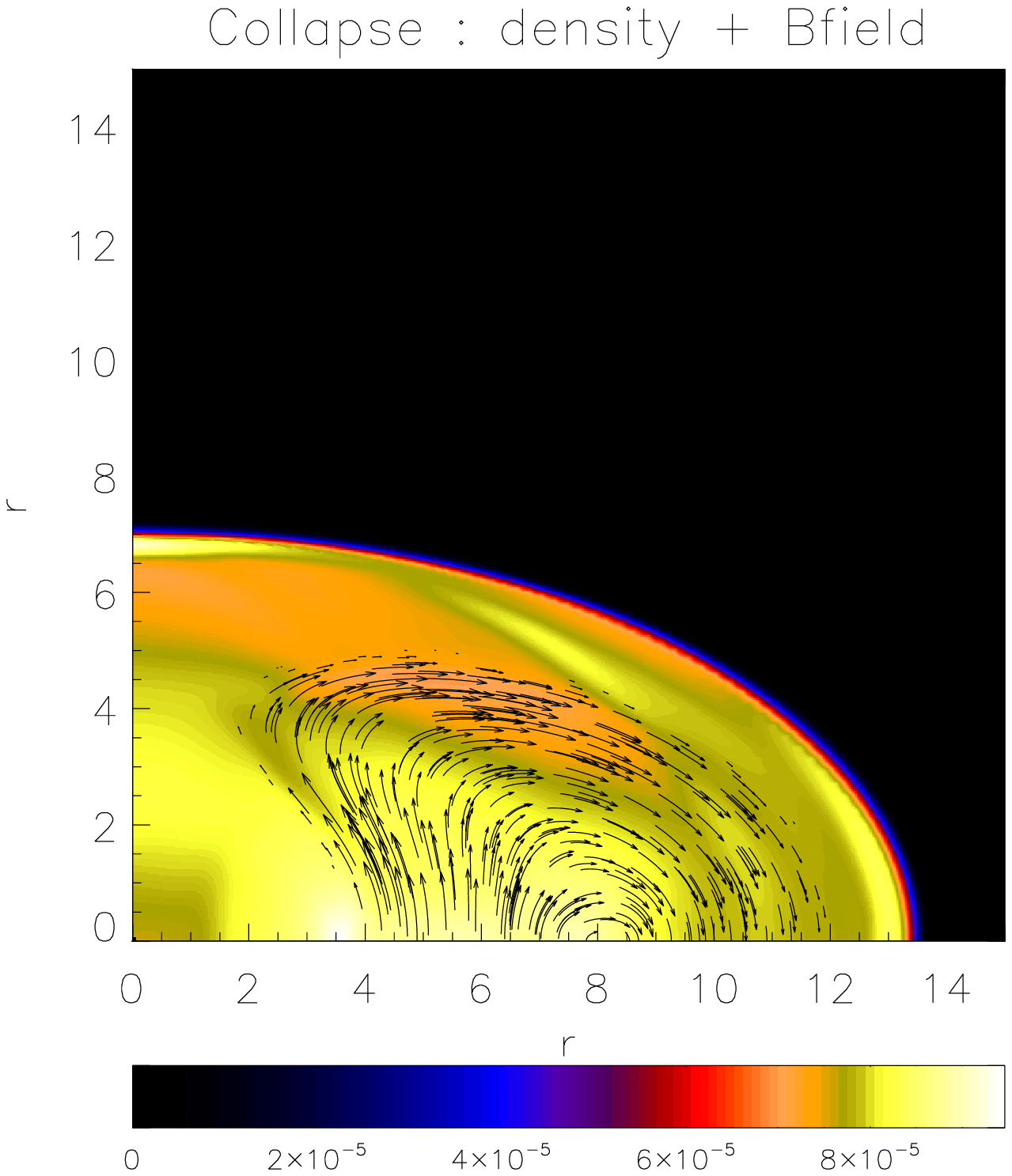}\includegraphics{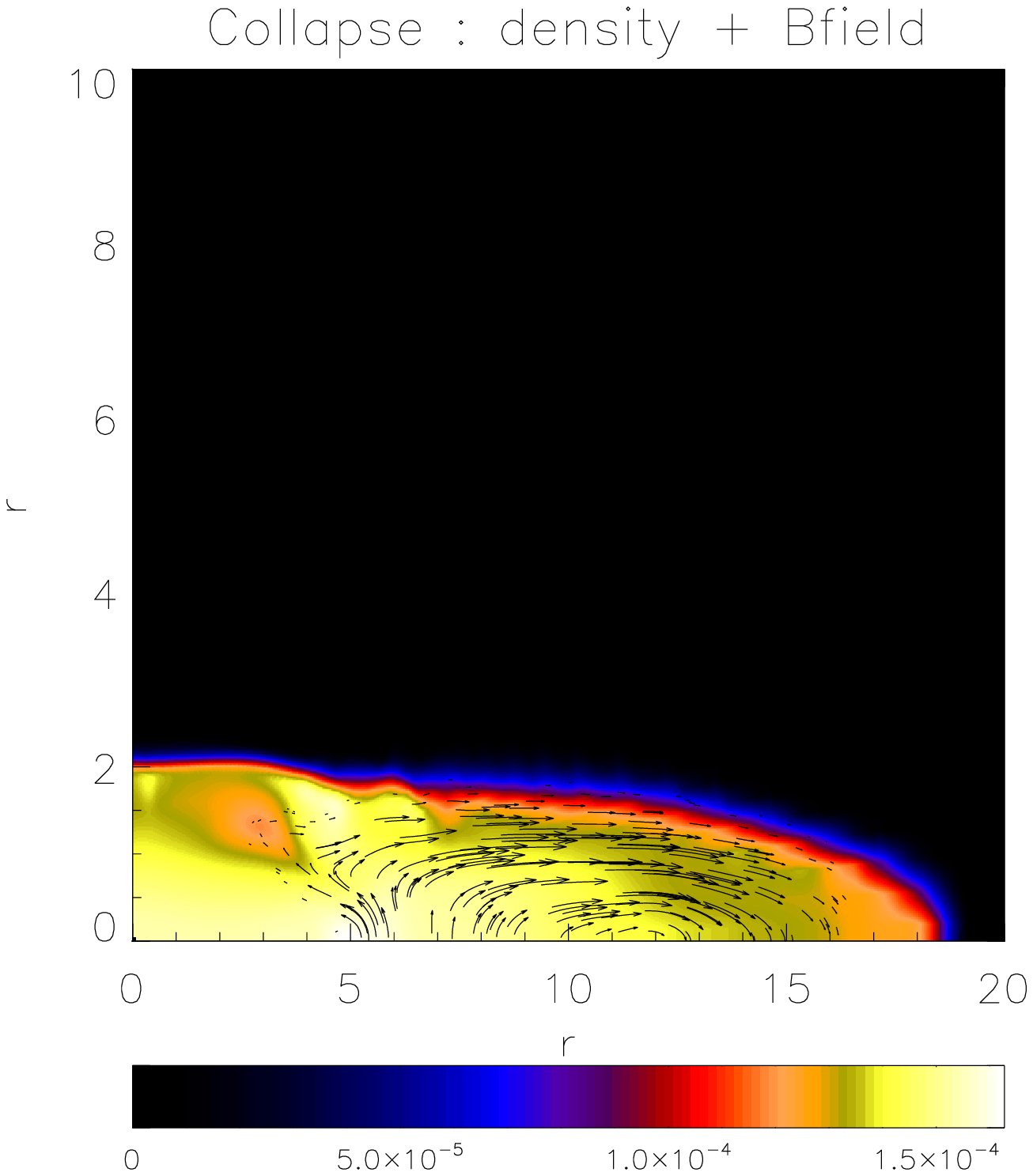}\includegraphics{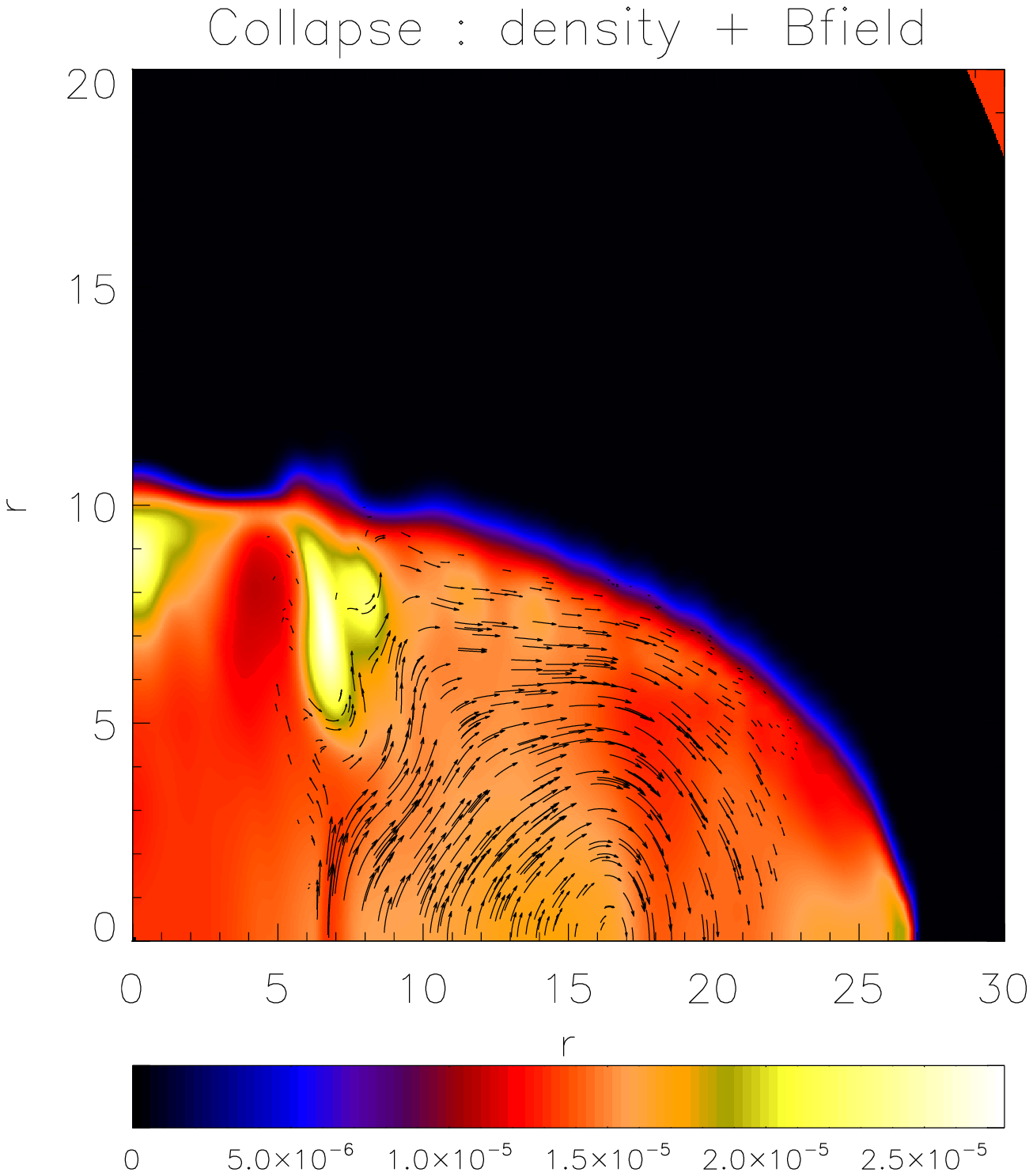}}
\caption{Evolution of the {\it toy collapse} model. Panels show the
  rest mass density and the mangetic field lines (represented by
  arrows) at three different instants of the evolution. Left panel:
  $t=50$, initial aspherical collapse. Note that the star is collapsing
 along the axis but expanding at the equator. Middle panel: $t=100$,
 bounce. The star has turned into a disk (aspect ration $\sim 1/10$). 
 Right panel: $t=200$, later expansion. The density is now about
 15\% of the initial value. A density bump is formed distorting the magnetic field. }
\label{fig:toycoll}
\end{figure*} 


In Fig.~\ref{fig:bhcollapse1} we show the evolution of a few
quantities. The apparent horizon forms at a time $t_{AH}\simeq 47$ to
be compared to $t_{AH}\simeq 48$ in \citet{bernuzzi10}. When the AH
forms, its location is $r=0.61$, the rest mass outside it is $0.28$ of the
initial rest mass,
and the value of the lapse at the center is $\alpha_c = 0.028$. Our results
agree with what is shown in Fig.~3 of \citet{cordero-carrion09}, and
Fig.~8 of \citet{bernuzzi10}. In Fig.~\ref{fig:bhcollapse1} we also
show the value of the central density. However a cautionary remark is
here in order about this quantity. As the collapse proceeds the metric
becomes progressively more curved in the origin and the density
correspondingly steeper, to the point that interpolation and round
off errors become more and more relevant. In \citet{cordero-carrion09},
despite their much higher resolution, already at $t \sim t_{AH}-10$
noise is present in the plot of the central density, which is no
longer monotonically increasing. In our case we find that, as the
system approaches the formation of the AH, the confidence and accuracy
with which we can derive the central density at $r=0$, by extrapolating the
values on the numerical grid, rapidly decreases. At the time the AH
forms we have evaluated that the extrapolation accuracy is such that
the possible error on the density is $\sim $ 10\%, and it rapidly
increases afterward. For this reason, we truncate the central density
plot at the formation of the AH. It should be reminded that the
standard practice to handle systems after the formation of an AH is to
excise the region inside the AH itself, excluding it from the
computational domain \citep{baiotti05, hawke05}. 
The precise value of the cental density is
obviously not an important quantity, as far as the global
evolution of the system is concerned. Indeed, quantities that depend on global
properties of the system like the AH radius, $r_{AH}$, the rest
mass left outside it, the value of the lapse at the center (which depends
on the global distribution of matter in the domain), all agree very well with
what has been previously found in the literature.

For the 2D case we consider the model D4 of \citet{baiotti05} and
\citet{cordero-carrion09}. This corresponds to a uniformly rotating neutron star with a
central density $\rho_c = 3.116 \times 10^{-3}$, a rotation rate
$\Omega = 0.0395$, a gravitational mass $M = 1.86$, an equatorial
radius $r_e = 7.6$, and an ellipticity $r_p/r_e=0.65$. Following
\citet{cordero-carrion09} the collapse is triggered by reducing the
pressure 2\% with respect to the equilibrium value.  The
computational domain $r=[0,10]$, $\theta =[0,\pi]$ is covered by 200
equally spaced radial zones, and 100 equally spaced angular zones. 
As usual, a hot low density atmosphere is set
outside the NS, and let evolve freely and an ideal gas EoS is used.
Again, despite our radial resolution being  50 times worse than the central
resolution used by \citet{cordero-carrion09}, we are able to follow the
evolution of the system  past the formation of an AH.

In Fig.~\ref{fig:bhcollapse1} we show the evolution of a few
quantities. The AH forms at a time $t_{AH}\simeq 126$, to
be compared to $t_{AH}\simeq 130$ in \citet{cordero-carrion09},
its location at that moment is $r = 0.75$, the rest mass outside 
is $23\%$ of the initial rest mass,
and the value of the lapse at the center is $\alpha_c = 0.025$. Our results
agree with what is shown in Fig.~3 of \citet{cordero-carrion09}. The
same considerations stated above for the central density still apply.
Interestingly, we are able to follow the evolution of the model D4 for a
much longer time after the formation of the AH with respect to
\citet{cordero-carrion09}. Fig.~\ref{fig:bhcollapse2} shows the
evolution of the NS as the AH forms and grows. The middle panel represents the
 system at the moment of the formation of the AH, as in their Fig.4
(warning: km units were used on the axes).
The lower panel shows the density at $t=t_{AH}+6$. Clearly a disk has
formed as in \citet{baiotti05}: the NS has been completely accreted
inside the AH in the polar region up to a latitude of around
$30^\circ$ while matter is still present in the equatorial region up
to $r\sim 3$.

\subsection{Toy collapse}
\label{sect:toycollapse}

Core-collapse simulations are often presented as a test for code
performances. However, they often aim at simulating realistic 
systems and involve complex physics: to the complexity of exact
MHD solutions in a dynamical metric \citep[][]{bocquet95}, they usually add 
sophisticated initial conditions from stellar evolution, tabulated EoS, and
neutrino transport. While all these elements are undoubtedly
important in the study of the physics of core collapse, it might be
questioned wether they are truly necessary to evaluate the performances of
the metric solver. Moreover, the diversity in setup, EoS, and transport
algorithms makes a direct comparison among different codes quite
difficult. Such simulations are also hard to reproduce, if, for example,
initial conditions or tabulated EoS are not easily and freely
available, or require specific code implementations. 
Moreover, results themselves often show the growth of turbulence
(convection) and MRI \citep{cerda-duran08}, for which comparison should be 
done in phase space, and not on selected snapshots at arbitrary times. 
In an attempt to construct a simple run with  fully defined initial
conditions, a simple EoS, and no complex physics, we have designed 
a novel test run of what we call a {\it toy collapse}.
However, this incorporates some of the important elements of a fully 2D 
non-linear evolution typical of a more realistic collapse scenario,
 and it also includes a poloidal magnetic field.

The chosen  initial conditions are the following
\be
\left\{ 
\begin{array}{l l}
  \rho=10^{-4} & \quad \mbox{}\\
  p=\mathrm{max}[\pi/3\, (10^2-r^2)\rho^2 , 10^{-9}] & \quad \mbox{if $r \le 10$}\\
  v^\phi=\alpha^{-1}(0.025+\beta^\phi)  & \quad \mbox{} \end{array} \right. 
\ee
\be
\left\{ 
\begin{array}{l l}
  \rho=10^{-7} & \quad \mbox{}\\
  p=10^{-9} & \quad \mbox{if $r > 10$}\\
  v^\phi=0.0 & \quad \mbox{} \end{array} \right. 
\ee
corresponding to a rigidly rotating uniform sphere. Note that the kinetic
pressure is $1/2$ of the hydrostatic value, for the same matter
distribution, in Newtonian gravity and with no rotation.
To this configuration a poloidal magnetic field is added using a generator
potential
\be
A_\phi = \mathrm{max}[3-\sqrt{[8(r-6)^2+36(\theta-\pi/2)^2]},0.],
\ee
yielding the components
\be
\mathcal{B}^r:=\gamma^{1/2}B^r = \partial_\theta A_\phi,
\ee
\be
\mathcal{B}^\theta:=\gamma^{1/2}B^\theta = \partial_r A_\phi,
\ee
where we recall that $\gamma^{1/2}=\psi^6 r^2\sin\theta$.

The initial conditions are found by solving the XCFC equations
together with the above conditions for the fluid and magnetic
variables. This provides us with a self-consistent set of conserved
variables and metric coefficients, corresponding to our choice
of primitive variables (note that our definition of velocity
implicitly involves the metric). The evolution of this configuration
incorporates various elements of strongly dynamical and aspherical
collapse. The original uniform sphere collapses preferentially along
the polar axis, where there is no centrifugal support, while it tends
to expand at the equator, where rotation is stronger, forming a
strongly oblate disk. Due to the stiff EoS, the system bounces when
the central density has increased of about 50\%, then
the structure re-inflates, driving a shock into the lower density of the
surrounding atmosphere. During the collapse the poloidal field is
stretched, compressed, and twisted by rotation, but remains confined
inside the star. In fig.~\ref{fig:toycoll} we show three snapshots of
the evolution, that we describe as follows: the quasi-spherical initial collapse,
the bounce, and the final structure. The numerical domain, $r=[0,35]$ and
$\theta=[0,\pi]$, is made up by 500 zones in the radial direction and by
150 zones in angle. The initial high density sphere is resolved over about
150 radial zones. The evolution is followed for a time
$t_\mathrm{max}=200$ corresponding in physical units to $\simeq 1$~ms.


\section{Conclusions}
\label{sect:concl}

In this paper we have upgraded the \emph{Eulerian conservative high-order} code
for GRMHD \citep[ECHO:][]{delzanna07} to dynamical spacetimes. We have chosen
a fully constrained method and conformal flatness for the Einstein equations,
and in particular we have built a numerical solver based on the 
\emph{extended conformally flat condition} scheme \citep[XCFC:][]{cordero-carrion09}
for the elliptic equations providing the metric terms. This is known
to improve on the previous CFC formulation \citep[e.g.][]{wilson03}, both
because of the hierarchical nature of the equations to solve (the elliptic
equations are fully decoupled), and because local uniqueness of the
solution is ensured even for highly dynamical non-linear cases.
Our novel scheme is here named X-ECHO and we also present
a code to produce self-consistent initial data (metric terms and GRMHD
quantities) for polytropic, differentially rotating, relativistic (neutron) stars 
with a toroidal magnetic field, named XNS. This code is
publicly available at
{\small \verb|http://sites.google.com/site/niccolobucciantini/xns|}
and we hope it will provide useful benchmark initial data for NS
evolution or core collapse in the magnetized case.

Both X-ECHO and XNS work in spherical coordinates of the conformally 
flat metric and axisymmetry is assumed. 
The 2D metric solver for the elliptic equations (Poisson-like scalar
PDEs and  vector Poisson PDEs) use a mixed technique: spectral
decomposition in spherical harmonics (or vector spherical harmonics)
in the angular direction and finite-differences leading to the inversion
of band-diagonal matrices in the radial direction. This is achieved
on the same numerical grid used for evolving the fluid/MHD quantities,
thus avoiding the need for interpolation over different meshes.
We fully test the codes against 
known problems involving fluid configurations in dynamical spacetimes,
basically 1D and 2D evolution and vibration modes for NS configurations,
migration to stable branches,  1D and 2D collapse of an unstable NS
towards a BH, including the formation of an apparent horizon,
and we propose a couple of novel GRMHD
test problems, the evolution of a differentially rotating magnetized NS and a 
\emph{toy collapse} simulation in the presence of poloidal magnetic fields.
The metric solver is fast, and, on the cases we have tested, the CPU
time required to solve the XCFC system, is comparable with the
time taken to update the MHD fluid quantities, despite the use of fast
Riemann solver (HLL/HLLC). For higher than second order reconstruction
tecniques, the computational time is always dominated by the HD/MHD
module. The code has been validated against previous results obtained
with both {\it free-evolution} and {\it fully costrained} schemes, and
with the linear theory for perturbations. Performances  in the
  presence of strong
magnetic fields, violently dynamical configurations, large
deviation from sphericity,  and  even apparent
horizons, show that the method and its implementation
with the HD/MHD module are stable in the situations of interest.
Moreover we have shown that, in the case of NSs surrounded by a low
density atmosphere, there is no need to apply any reset
procedure to the atmosphere itself, which can be let evolve freely,
without altering the stability or the results of the simulations.

For the immediate future we plan to investigate the stability and
to find the characteristic vibrational modes of a set of magnetized 
NS configurations, with and without differential rotation, both for
stable and unstable megnetic profiles. We also plan to investigate the
growth of magnetic field due to MRI in differentially rotating NS,
which might be of some interest to explain the late flaring activity
that is observed in long duration GRBs, and that, within the
millisecond-magnetar model, is commonly attributed to bursty magnetic
activity in the cooled and convectively stable NS.
However, the final goal is to include a more realistic treatment of
the microphysics, going beyond the simple ideal gas law implemented
here for reproducibility of the numerical tests, especially as far
as neutrino heating is concerned, and possibly to couple our code with
a transport algorithm for neutrinos as required for collapse calculations.
This will allow us to study in details the magnetized core-collapse scenario,
and to investigate the role of a strong magnetic field in shaping and regulating 
the collapse. More in particular, we also plan to derive from magnetized
collapse simulations a more realistic setup for the (long) GRB model 
recently proposed by \cite{bucciantini09}, where a 
newly born millisecond proto-magnetar drives a GRMHD wind, that, 
due to confinement of the external stellar envelopes, 
collimates relativistic jets escaping the progenitor along the poles. 


\begin{acknowledgements}
We sincerely thank Sebastiano Bernuzzi for having introduced us
to the world of the Einstein equations and for suggesting us some
useful tests and benchmarks, and Luca Franci for
practical help with the use of the RNS code for initial data. We also
thanks the anonymous referee for her/his helpful suggestions.
The work of NB has been supported by a NORDITA fellowship grant.
\end{acknowledgements}   
  

\bibliographystyle{aa}
\bibliography{grmhd}



\end{document}